\documentclass[manuscript]{aastex}
\usepackage{graphicx}
\usepackage{url}

\shorttitle{A complete grid of reflection models}
\shortauthors{Garc\'{\i}a \& et al.}

\begin{document}

\title{X-ray reflected spectra from accretion disk models. \\
III. A complete grid of ionized reflection calculations}

\author{J. Garc\'ia}
\affil{Harvard-Smithsonian Center for Astrophysics, 60 Garden St., Cambridge, MA 02138 USA \&
Department of Astronomy, University of Maryland, College Park, MD, USA}
\email{javier@head.cfa.harvard.edu}

\author{T. Dauser}
\affil{Dr. Karl Remeis-Observatory and Erlangen Centre for Astroparticle Physics, Sternwartstr. 7, 96049 Bamberg, Germany}
\email{thomas.dauser@sternwarte.uni-erlangen.de}

\author{C.~S. Reynolds}
\affil{Department of Astronomy, University of Maryland, College Park, MD, USA
\& Joint Space-Science Institute, University of Maryland, College Park, MD, USA}
\email{chris@astro.umd.edu}

\author{T.~R. Kallman}
\affil{NASA Goddard Space Flight Center, Greenbelt, MD 20771}
\email{timothy.r.kallman@nasa.gov}

\author{J.~E. McClintock}
\affil{Harvard-Smithsonian Center for Astrophysics, 60 Garden St., Cambridge, MA 02138 USA}
\email{jem@cfa.harvard.edu}

\author{J. Wilms}
\affil{Dr. Karl Remeis-Observatory and Erlangen Centre for Astroparticle Physics, Sternwartstr. 7, 96049 Bamberg, Germany}
\email{joern.wilms@sternwarte.uni-erlangen.de}

\author{W. Eikmann}
\affil{Dr. Karl Remeis-Observatory and Erlangen Centre for Astroparticle Physics, Sternwartstr. 7, 96049 Bamberg, Germany}
\email{wiebke.eikmann@sternwarte.uni-erlangen.de}

%
\begin{abstract}

We present a new and complete library of synthetic spectra for
modeling the component of emission that is reflected from an
illuminated accretion disk.  The spectra were computed using an
updated version of our code {\sc xillver} that incorporates new
routines and a richer atomic data base.  We offer in the form of a
table model an extensive grid of reflection models that cover a wide
range of parameters.  Each individual model is characterized by the
photon index $\Gamma$ of the illuminating radiation, the ionization
parameter $\xi$ at the surface of the disk (i.e., the ratio of the
X-ray flux to the gas density), and the iron abundance $A_\mathrm{Fe}$
relative to the solar value.  The ranges of the parameters covered
are: $1.2 \leq \Gamma \leq 3.4$, $1 \leq \xi \leq 10^4$, and $0.5 \leq
A_\mathrm{Fe} \leq 10$.  These ranges capture the physical conditions
typically inferred from observations of active galactic nuclei, and
also stellar-mass black holes in the hard state.  This library is
intended for use when the thermal disk flux is faint compared to the
incident power-law flux.  The models are expected to provide an
accurate description of the Fe K emission line, which is the crucial
spectral feature used to measure black hole spin.  A total of 720
reflection spectra are provided in a single FITS 
file\footnote{\url{http://hea-www.cfa.harvard.edu/~javier/xillver/}} suitable for the
analysis of X-ray observations via the {\tt atable} model in {\sc
xspec}.  Detailed comparisons with previous reflection models
illustrate the improvements incorporated in this version of {\sc
xillver}.

\end{abstract}
%
%
\section{Introduction}
 A large number of observations in the past few decades have established the
general properties of the X-ray spectra from accreting luminous sources. These
are systems in which a compact object, either a neutron star or a black hole, affects
the surrounding material with its vast gravitational forces, which leads to accretion
of large quantities of gas and consequently to the release of large amounts of
radiation, mostly in the X-ray band. The compact object can be a supermassive black hole
as those present in the active galactic nuclei (AGN) of many galaxies, or of
stellar-mass, such as is the case of many galactic black holes (GBHs).

In AGN, the X-ray continuum is usually characterized by a power-law spectrum
with photon index $\Gamma$, which can typically be observed to lie in the 
$1.8 \gtrsim \Gamma \gtrsim 2.2$ range, although in some extreme cases it can 
be either closer to 1 or larger than 3, extending to high-energies with an 
exponential cutoff around $100-300$~keV.
At energies below $0.1$~keV, the power-law often meets a soft excess 
that mimics a blackbody radiation. In GBHs, the X-ray spectrum in most cases
is dominated by a thermal blackbody-like component that peaks at high-energies
($0.2 \lesssim kT \lesssim 2$~keV), but can also show a high-energy tail component
of emission. In general, this spectral energy distribution is explained in terms of a 
geometrically thin, optically-thick accretion disk around the compact object. 
The energy dissipation within the disk would be responsible for the quasi-blackbody 
emission observed. The power-law continuum is believed to originate through the Compton 
up-scattering of the thermal photons by the electrons in a hot corona or a jet
\citep{haa93,dove97,dau13}. 

The presence of this dense ($n_{\mathrm H} \gtrsim 10^{12}$~cm$^{-3}$), warm ($T\sim 10^5-10^7$~K), 
and optically-thick ($\tau_{\mathrm T} \gtrsim 1$) medium is also supported by the detection of atomic 
features from several ions. These and other features constitute an important component of the
X-ray spectrum observed from accreting sources, resulting from the reprocessing of radiation by
the material in the disk. This component is commonly referred to as {\it reflection}, in the sense that
it is the result of radiation that is returned from the accretion disk by fluorescence or
electron scattering. The current paradigm is that the original power-law radiation irradiates
the surface of the accretion disk. The X-ray photons then interact with the material producing
diverse atomic features. These can be produced both via absorption (mostly in form of edges), 
and emission (in form of fluorescence lines and radiative recombination continua, RRC).
Therefore, the reflection component provides direct information about structure, temperature, 
ionization stage, and composition of the gas in the accretion disk.

The presence of the Fe K-shell fluorescence emission and the absorption K-edge observed
in the $6-8$~keV energy range are recognized as strong evidence for reflection. X-ray photons
that are photoelectrically absorbed have enough energy to remove a 1s electron from its K-shell,
leaving it in a quasi-bound state above the continuum (autoionizing state). The K-hole is 
then filled by an electron, and the energy difference can be released by emitting a second
electron (Auger process), or by the emission of a K-shell photon. These transitions are of the 
parity-changing type $np-1s$, where $n=2$ and $n=3$ correspond to the K$\alpha$
and K$\beta$ transitions, respectively. Higher $n$ transitions are allowed but with a much 
lower probability. The fluorescence yield,
i.e. the probability of emitting a photon over an Auger electron, is proportional to the nuclear
charge $Z$ to the forth-power (i.e., $\propto Z^4$), making the Fe K-shell emission particularly
strong. This has been shown to be true in a large number of observations from X-ray accreting
sources \citep[e.g.][]{got95,win09,ng10}. The Fe K emission line has also proven to be crucial in 
the determination of one of the fundamental quantities that describe black holes, its angular
momentum \citep{lao91,dab97,bre06}. If the reflection occurs near the black hole, line photons will suffer
Doppler effects, light bending, and gravitational redshift, which produces a skewed line profile
with a red wing that can extend to very low energies, particularly in the case of high spin
\citep[e.g.,][]{fab00,fab03,rey03,dov04,mil08,ste11,rey12,dau12}. 
Therefore, the proper modeling of the reflected disk 
component is of vital importance for determining one of the two fundamental parameters that
define a black hole.

The first theoretical studies of X-ray reflection assumed that irradiation on 
the surface of the accretion disk was weak enough so the gas remains neutral, but yet would
reprocess the radiation producing observable spectral features \citep{gui88}. Green's functions
to describe the scattering of photons by cold electrons were first derived by \cite{lig80,lig81},
and their implications for AGN observations discussed in \cite{lig88}. The latter approach
is used for the calculation of cold reflection in the model {\sc pexrav} \citep{mag95}. However,
none of these calculations included line production. \cite{geo91} and \cite{mat91} included
the X-ray fluorescence line emission in their Monte Carlo calculations, providing line strength,
angular distribution, and equivalent widths for the Fe K line. \cite{zyc94} carried out similar
calculations including photoionization equilibrium, yet neglecting the intrinsic emission
inside the gas.

Much more detailed calculations of the radiative transfer of X-rays in an optically-thick 
medium were carried out by \cite{ros78} and \cite{ros79}. Their code solves the transfer of the continuum
photons using the Fokker-Planck diffusion equation, including a modified Kompaneets operator
to properly treat the Compton scattering, while the transfer of lines is calculated using the
escape probabilities approximation. This code has been updated over the years leading
to the {\sc reflionx} model \citep{ros93,ros05}, which has been widely used to model the
reflected component in the spectra of accreting X-ray sources.
\cite{dum03} used the {\sc titan} code to examine the accuracy
of the escape probability methods versus the ``exact" solution of the radiative transfer by
implementing accelerated lambda iterations. This code has been extended by \cite{roz02} to treat the cases
of Compton-thick media. All these calculations assume constant density in the material. It
has been argued that a plane-parallel slab under hydrostatic equilibrium could represent the surface
of an accretion disk more accurately \citep{nay00}, and that its reflected spectrum is in fact 
different from the one predicted by constant density models \citep[see also][]{roz96,nay01,
bal01,dum02,ros07,roz08}.

Besides the techniques used to treat the transfer of photons through the media, the codes 
mentioned above may also differ in terms of the atomic data used, which in most cases 
offers a limited perspective of the physics governing the atomic processes for the absorption,
excitation, and emission processes. These limitations have been overcome by our new reflection 
model {\sc xillver} \citep{gar10,gar11}.
{\sc xillver} calculates the reflected spectrum emerging from the surface of an X-ray illuminated
accretion disk by simultaneously solving the equations of radiative transfer, energy balance, 
and ionization equilibrium in a Compton-thick, plane-parallel medium. The transfer is solved
using the Feautrier method under a lambda iteration procedure \citep{mih78}. {\sc xillver} implements 
the well-known photoionization code {\sc xstar} \citep{kal01} to solve the ionization structure of
the atmosphere, therefore making use of the most updated, accurate, and complete atomic database 
for the X-ray line emission, photoabsorption, and photoionization for all the astrophysically
relevant ions.

In this paper we present a complete library of reflection spectra using an updated version
of our code {\sc xillver}. This grid of models covers a wide range of input parameters, relevant
to model the spectrum from accreting X-ray sources. Each model is defined by the photon index
$\Gamma$, the ionization parameter $\xi$ (given by the ratio of the X-ray flux over the gas 
density), and the abundance of Fe with respect to its solar value. The models are provided in
a single table\footnote{\url{http://hea-www.cfa.harvard.edu/~javier/xillver/}}
suitable to be used in the commonly used fitting packages such as {\sc xspec}
\citep{arn96}, and {\sc isis} \citep{hou00}. 
We show a detailed analysis of our models along the space of parameters and a 
careful comparison with other similar models. The improvements achieved are stressed, and their 
implications on the analysis of X-ray spectra is discussed.

This paper is organized as follows. In Section~\ref{secmod} we describe the basic aspects of 
the theory used in our calculations, paying particular attention to the changes and improvements
implemented in the code. An analysis of the new reflection models for the different parameters 
is presented in Section~\ref{secres}. We show the effect of the photon index, the
ionization parameter, and the iron abundance on the ionization structure and ultimately on
the emergent spectra. A thorough comparison with other reflection models is also provided. 
The main conclusions and future prospects are summarized in Section~\ref{seccon}.

%
%
\section{The Reflection Model}\label{secmod}
In order to calculate the reflected spectra from X-ray illuminated 
accretion disks we made use of our reflection code {\sc xillver}. 
The details of the calculations are fully described in \cite{gar10}, 
thus we shall review only the main aspects, in particular those where 
some changes have been applied. One important modification takes place in the
solution of the radiation transfer equation. This equation describes the interaction of the
radiation field with the gas in the illuminated atmosphere, and it is now
expressed as
\begin{equation}\label{ert}
\mu^2\frac{\partial^2 u(\mu,E,\tau)}{\partial\tau^2}
= u(\mu,E,\tau) - S(E,\tau)
\end{equation}
where $u(\mu,E,\tau)$ is the energy density of the radiation field for 
a given cosine of the incident angle with respect to the normal $\mu$, energy $E$, 
and position in the slab, which now is specified in terms of the {\it total}
optical depth 
\begin{equation}\label{eqdt}
d\tau =\chi(E) dz,
\end{equation}
where 
\begin{equation}
\chi(E)=\alpha_{kn}(E) + \alpha_a(E)
\end{equation}
is the total opacity of the gas that includes both scattering and absorption.
Here, $\alpha_{kn}(E)$ and $\alpha_a(E)$ are the scattering and absorption coefficients,
respectively. The former is given by the product of the gas density times the Klein-Nishina 
cross section $\sigma_{kn}(E)$ for electron scattering (which includes the relativistic
correction). The latter is defined by the product of the gas density times the absorption cross section
$\sigma_a(E)$ due to bound-bound, bound-free, and free-free processes.
The source function $S(E,\tau)$ in Equation~(\ref{ert}), is given by
the ratio of the total emissivity to the total opacity.
It can written as
\begin{equation}\label{eqsou}
S(E,\tau) = \frac{\alpha_{kn}(E)}{\chi(E,\tau)}J_c(E,\tau) + \frac{j(E,\tau)}{\chi(E,\tau)},
\end{equation}
where $j(E,\tau)$ is
the continuum plus lines emissivity, and $J_c(E,\tau)$ is the Comptonized mean
intensity of the radiation field resulting from the convolution
\begin{equation}\label{eqjc}
J_c(E,\tau) = \frac{1}{\sigma_{kn}(E)}\int dE' J(E',\tau)\sigma_{kn}(E')P(E',E).
\end{equation}
Here, $J(E,\tau)=\int u(\mu,E,\tau)d\mu$ is 
the unscattered mean intensity of the radiation field. The quantity $P(E',E)$ is the
probability of a photon with energy $E$ to be Compton scattered to an energy $E'$,
which we approximate by assuming a Gaussian profile centered at 
\begin{equation}\label{eqec}
E_c = E \left( 1 + \frac{4kT}{m_ec^2} - \frac{E}{m_ec^2}\right)
\end{equation}
\citep{ros78,ros93}, with energy dispersion
\begin{equation}\label{eqsig}
\sigma = E \left[ \frac{2kT}{m_ec^2} + \frac{2}{5}\left(\frac{E}{m_ec^2}\right)^2\right]^{1/2},
\end{equation}
where $k$ is the Boltzmann's constant, $T$ is the gas temperature, $m_e$ is the rest-mass
of the electron, and $c$ is the speed of light. In Equation~(\ref{eqjc}) we now follow 
\cite{nay00} and properly take into account the relativistic (Klein-Nishina) 
correction to the cross section for electron scattering, rather than using the classical (Thomson)
approximation \citep{gar10}.

Referring to Section~2 in \cite{gar10}, the radiative 
transfer equation written in terms of the Thomson optical depth $\tau_{\mathrm T}$ has
an extra linear term when compared with the usual form shown here (Equation~\ref{ert}).
We found that the radiation transfer equation written in terms of the total
optical depth behaves better when there are large gradients in the opacity, despite
the fact that the integration along the path is performed in a different grid for each
energy bin (i.e., the total optical depth is a function of the photon energy). We use
a logarithmically spaced grid fixed in $\tau_{\mathrm T}$ with 200 spatial bins in the 
$10^{-4} \leq \tau_{\mathrm T} \leq 10$ range. At each energy, the
total optical depth is calculated by integrating Equation~(\ref{eqdt}).

The boundary conditions are only slightly different from those in \cite{gar10}. At the top of 
the slab ($\tau=0$), we specify the radiation field incident at a given angle 
$\mu_0$ by
\begin{equation}\label{eqbco}
\mu \left[ \frac{\partial u(\tau,\mu,E)}{\partial \tau} \right]_0
- u(0,\mu,E) = -\frac{2F_x}{\mu_0}\delta(\mu-\mu_0),
\end{equation}
where $F_x$ is the net flux of the illuminating radiation integrated in the whole energy
band. At the inner boundary ($\tau=\tau_{max}$), we specify the outgoing radiation field 
to be equal to a blackbody with the expected temperature for the disk:
\begin{equation}\label{eqbcm}
\mu \left[ \frac{\partial u(\tau,\mu,E)}{\partial \tau} \right]_{\tau_{max}}
+ u(\tau_{max},\mu,E) = B(T_{disk}),
\end{equation}
where $B(T)$ is the Planck function, and $T_{disk}$ can be defined using the 
\cite{sak73} formulae. Nevertheless, for the models presented in this paper we have
chosen to neglect any illumination from below the atmosphere, i.e., $B(T)=0$. This
is convenient when comparing our models with previous reflection calculations such
as {\sc reflionx} \citep{ros05}. Also, for the parameters typical
in AGN, the temperature of the disk is sufficiently low that the intrinsic black body
emission is weak compared to the power-law incident at the surface.
Since these models are calculated under the assumption of 
constant density, we use the common definition of the ionization parameter \citep{tar69} 
to characterize each case, namely,
\begin{equation}\label{eqxi}
\xi = \frac{4\pi F_{\rm x}}{n_e},
\end{equation}
where $F_{\rm x}$ is the net integrated flux in the 1-1000~Ry energy range, and 
$n_e=1.2 n_{\mathrm H}$, and $n_e$ and $n_{\mathrm H}$ are the the electron 
and hydrogen number densities, respectively. The solution of the system is found by 
forward elimination and back substitution. With a solution for the radiation field
$u(\mu,E,\tau)$, a new $J_c(E,\tau)$ and thus a new $S(E,\tau)$ can be calculated.
These are then used to update the solution for $u(\mu,E,\tau)$.
The procedure is repeated until these quantities stop changing within a small fraction.
A full transfer solution must be achieved iteratively in order to self-consistently 
treat the scattering process. In general, this procedure 
requires $\sim \tau_{max}^2$ iterations for convergence. Since our calculations are 
carried out up to $10$ Thomson depths, we perform 100 iterations. This ensures convergence
for energies where scattering is dominant, for which $\tau_{\mathrm T} \approx \tau$. For
energies where the photoelectric opacity is large, the total optical depth can be
much larger than 10. Nevertheless, if absorption dominates over scattering, the first
term in the right-hand side of Equation~(\ref{eqsou}) is reduced. In the limit 
$\alpha_{a} \gg \alpha_{kn}$, the source function essentially becomes independent of 
$J_c(E,\tau)$, and the solution converges very rapidly.

The structure of the gas is determined by solving the ionization balance equations
for a given gas density and for a particular solution of the radiation field. At each point
within the slab, we use the photoionization code {\sc xstar} \citep{kal01} to calculate
level populations, temperature, the total opacity $\chi(E,\tau)$, and the emissivity 
$j(E,\tau)$, assuming that all the physical processes are in steady-state and imposing 
radiative equilibrium. We have updated our code {\sc xillver} to work with the current 
version of {\sc xstar} (version 2.2.1bn)
\footnote{\url{http://heasarc.gsfc.nasa.gov/xstar/xstar.html}},
and the most recent atomic database. Although the physics is the same, this new version 
includes improved routines that considerably reduce the computing time, allowing the 
exploration of a wider parameter space for a reasonable allocation of resources.

\subsection{Atomic Data}\label{secatomicd}

As mentioned before, {\sc xillver} implements the {\sc xstar} routines for the calculation
of the ionization structure of the gas, while making use of {\sc xstar}'s atomic data. The core of
the {\sc xstar} atomic database is described in detail in \cite{bau01}. It has been 
constructed using information from many sources such as  CHIANTI \citep{lan06}, ADAS 
\citep{sum04}, NIST \citep{ral08}, TOPbase \citep{cun93} and the IRON project \citep{hum93}. 
Over the last several years, we have dedicated a significant effort towards the investigation
of the K-vacancy states in many ions. This has allowed us to systematically improve the modeling 
of the K lines and edges relevant for high-quality astronomical X-ray spectra. 

Such atomic data sets include energy levels, wavelengths, Einstein $A$-coefficients, 
radiative and Auger widths computed for a large number of ions using three
different atomic-structure theoretical approaches, including the relativistic
Hartree--Fock \citep{cow81}, the {\sc autostructure} \citep{bad86,bad97}, 
and the multi-configuration Dirac--Fock \citep{ber87,sea87} methods. 
Extensive calculations of photoabsorption and photoionization cross sections have 
been also performed using the Breit--Pauli R-Matrix method including the effects of 
radiative and Auger damping by means of an optical potential \citep{gor99,gor00}.

In particular, calculations of the atomic data required for the spectral
modeling of the K-shell photoabsorption of oxygen and nitrogen ions have been carried
out in \cite{gar05} and \cite{gar09}. Computations have also been carried out for the
atomic structure of the isonuclear sequences of Ne, Mg, Si, S, Ar, and Ca \citep{pal08a},
nickel \citep{pal08b}, and aluminum \citep{pal11}. More recently, we have focused our
attention on the iron peak and light odd-Z elements, i.e., F, Na, P, Cl,
K, Sc, Ti, V, Cr, Mn, Co, Cu and Zn producing the atomic parameters for more than 3 million
fine-structure K lines \citep{pal12}. Photoabsorption calculations of the cross sections
across the K-edge of Ne, Mg, Si, S, Ar, and Ca have been performed for ions with less
than 11 electrons by \cite{wit09}, and for ions with more than 10 electron by \cite{wit11}.
In these calculations it was shown that damping processes affect the resonances 
converging to the K thresholds
causing them to display symmetric profiles of constant width that smear the
otherwise sharp edge at the photoionization thresholds. The Li-like to Ca-like
ion stages of nickel are discussed in \cite{wit11b}. These new data sets are continually
being incorporated into the {\sc xstar} data base  in order to generate improved opacities 
in the K-edge regions of the ions considered.

The main differences in the current version of the atomic data with respect to the one used in 
\cite{gar10} are as follows: The inclusion of the atomic data for the $n=2\rightarrow 3$
inner-shell transitions of Fe~{\sc vi-xvi} ions, the so-called Fe M-shell unresolved transition
array \citep{gu06}. The inclusion of the K-shell atomic data for the Mg, Si, S, Ar, Ca, Al, and Ni 
isonuclear sequences \citep{pal08a,pal08b,wit11,pal11,wit11b}. The implementation of the
updated radiative and dielectronic recombination rates from \cite{bad06} and \cite{bad03}.
Also, we have improved the resolution of the high-energy extrapolation implemented in the Fe 
photoionization cross sections for energies well above the K-edge. 
Finally, by looking at the third-row Fe ions (Fe~{\sc i-viii}), we found that 
for the densities used in our simulations ($n_e\sim 10^{15}$~cm$^{-3}$), many metastable
states are populated. Previously, we only used K-shell photoionization data for the ground
state, therefore missing the contribution from these metastable states. This resulted 
in a weak Fe K emission line in models with low ionization parameter. This problem has been
corrected in the new version of the data base by including the photoionization cross 
sections for all these states.

\subsection{Illumination spectrum}

\begin{figure}
\epsscale{1.0}\plotone{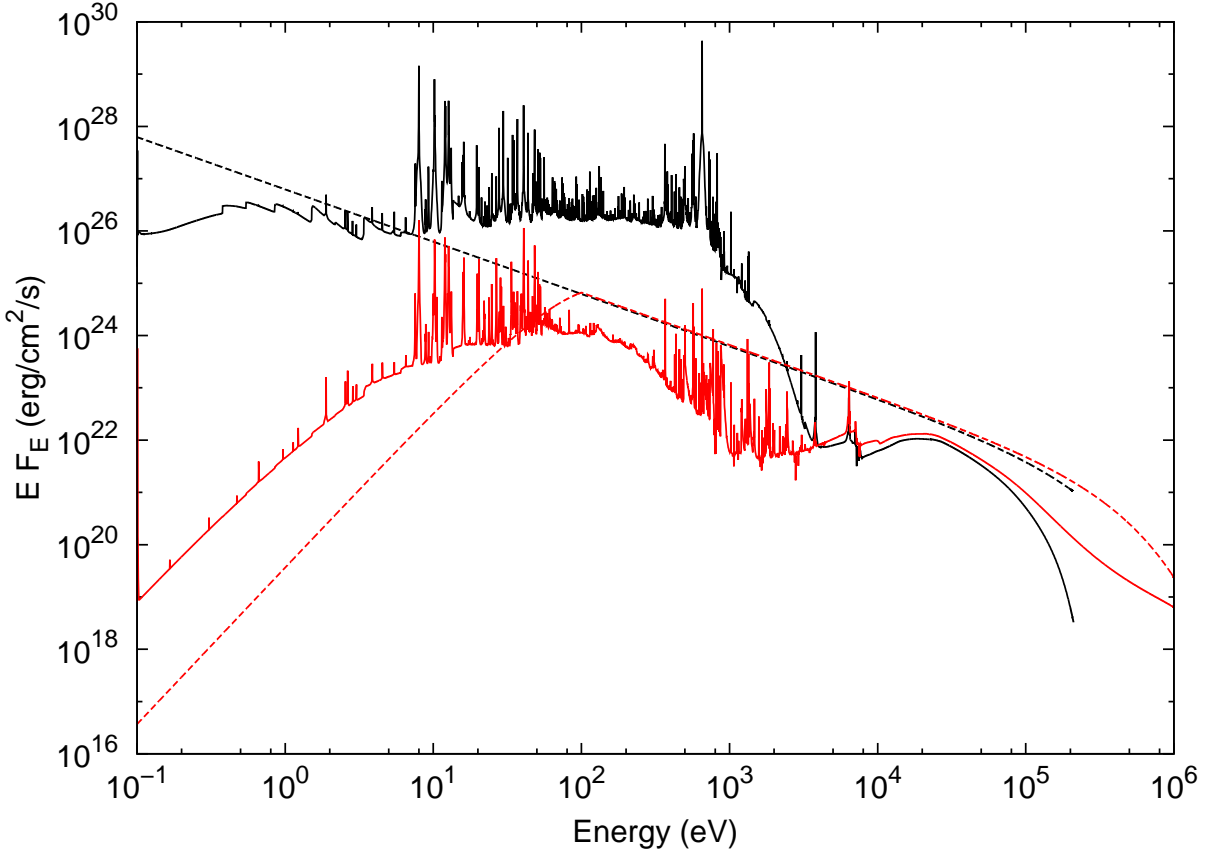}
\caption{Results using different illumination spectra with {\sc xillver} for a gas
with log~$\xi=2$. The dashed lines are the two different illumination power-law with
$\Gamma=3$. In black, the power-law covers the $0.1-2\times10^5$~eV energy range, with
an exponential cutoff at $200$~keV. In red, the power-law has a sudden cutoff at $0.1$~keV,
extends to $1$~MeV, and an exponential cutoff at $300$~keV. The black and red solid
lines are the resulting reflected spectra correspondent to each case.}
\label{fpowerlaw}
\end{figure}

Another important change in the {\sc xillver} code is the definition of the radiation field
that illuminates the surface of the accretion disk. In \cite{gar10}, the incident spectrum was
assumed to be a power-law with a given photon index $\Gamma$ for all energies in the 
$10^{-1}-2\times 10^5$~eV range. This kind of illumination is appropriate for spectra with 
$\Gamma\sim 2$. However, for steeper spectra ($\Gamma > 2$) this choice of illumination creates
unphysical conditions, since the incident power-law will tend to have very many
photons concentrated in the low energy part of the spectrum. Thus, we have now adopted a 
power-law that breaks at 100~eV and decays with an exponential tail for lower energies. We have
also extended the energy to 1~MeV, and moved the exponential high-energy cutoff from 200~keV
to 300~keV. Figure~\ref{fpowerlaw} shows the impact of this modification for the resulting reflected
spectrum in a case where $\Gamma=3$. The solid lines show the reflected spectra predicted by {\sc xillver} using the old
(black) and the new (red) definitions of the illumination. The incident power-law is shown with
dashed lines for each case. Both calculations have a similar ionization parameter (log~$\xi\sim 2$)
but the integrated flux is different given the differences in the shape of the power-law. Thus,
the model with the new definition (broken power-law) needed to be scaled down such that both incident
spectra will have the same flux near $100$~eV and energies above, region in which the 
photon index is $\Gamma=3$. 
The huge impact of the illuminating spectrum on the reflected component is clear, especially 
at energies below 5~keV.
This difference becomes more dramatic as the photon index increases. However, for harder
spectra ($\Gamma \lesssim 2$), the break in the incident power-law has little impact on the 
reflected spectrum, especially in the high-energy band.

\begin{figure}
\epsscale{1.0}\plotone{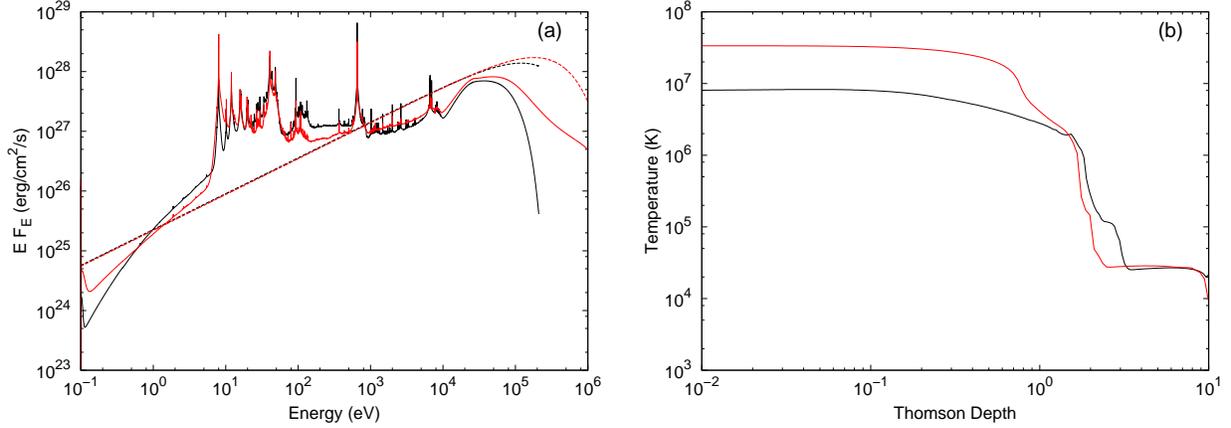}
\caption{Results using different illumination spectra with {\sc xillver} for a gas
with log~$\xi=2.8$. In panel (a), the dashed lines are the two different illumination power-law with
$\Gamma=1.4$, only differing by their high-energy limit. In black, the power-law covers 
up to $2\times10^5$~eV with an exponential cutoff at $200$~keV. In red, the power-law 
extends to $1$~MeV with an exponential cutoff at $300$~keV. The black and red solid
lines are the resulting reflected spectra correspondent to each case. Panel (b) shows the
respective temperature profiles within the slab.}
\label{fhighenergy}
\end{figure}

The extension of the spectrum to higher energies has also an important influence on
the reflection calculations. This is particularly important in models on the other extreme of the
photon index parameter space. In the case of a very hard illumination spectrum ($\Gamma < 2$), models 
with similar ionization parameters will differ in the number of photons that are concentrated in 
the high energy part of the spectrum. Figure~\ref{fhighenergy} shows the resulting {\sc xillver} models
for an incident power-law with $\Gamma=1.4$, log~$\xi=2.8$, and solar abundances. For the black 
and red curves, the energy range extends upward to $200$~keV and $1$~MeV, respectively.
In panel~(a), the 
dashed lines show the incident power-law spectra, while solid lines are the reflected spectra for
each case. The emission lines in the model with the high energy coverage (red) are considerably weaker
than those in the original model (black). This is most evident for the Fe K complex around $6-7$~keV,
and for the O~{\sc vii} L$\alpha$ near 650~eV. The reason for this difference is that in models with the 
high-energy extension, the region of the atmosphere near the surface is more strongly heated
by the high-energy photons, and thus its temperature is higher than in the original
model. This can be seen in panel~(b) of Figure~\ref{fhighenergy}, where the temperature profiles
in the vertical direction of the disk are plotted as function of the Thomson optical depth. Clearly,
the two models converge at large depths ($\tau_{\mathrm T}\sim$few), but the one in which the 
illumination extends to higher energies results in a considerably hotter atmosphere near the surface.
Although we expect the Fe features to be produced well inside the disk, line photons that escape
through a hotter gas will tend to suffer a larger energy shift from Compton scattering, 
with similar probabilities
for up- or down-scattering. This makes the line profile broader and more symmetric.
%
%
\section{Results}\label{secres}
As in the previous version of {\sc xillver}, all the models shown here are calculated with an energy
spectral resolution comparable with the best current observations (${\cal R}=E/\Delta E\sim 350$),
which requires 5000 bin points in the energy range considered here ($10^{-1}-10^{6}$~eV). Each
model covers a large column density of gas ($\tau_{\mathrm T}=10$), using 200 spatial zones and
10 bins that describe the angular dependence of the radiation fields. The first step in the 
spatial grid has been reduced to $\tau_{\mathrm T}\sim 10^{-4}$ to ensure an accurate representation of 
the illuminated layers in the slab. The hydrogen number density is held constant at 
$n_{\mathrm H}=10^{15}$~cm$^{-3}$ for all the models presented in this paper.
\subsection{The space of parameters}

A new library of reflection spectra has been produced using the new version
of the code {\sc xillver}. This set of models covers a wide range 
of parameter values that are relevant for fitting the spectra
of accreting sources. Each model is characterized by the photon index $\Gamma$
of the incident power-law radiation, the ionization parameter $\xi$, and the 
iron abundance $A_\mathrm{Fe}$ with respect to the solar value \citep{gre98}. In order
to have a library of models suitable for both AGN and GBHs, we have produced models
covering photon indices in the $\Gamma = 1.2 - 3.4$ range, in steps of 0.2.
Observations typically show that the reflection signatures from accretion disks
in AGN are produced in a material at a lower ionization stage than those observed
from GBHs \citep{gar11}, consistent with the picture that the accretion disk in the latter are 
much hotter than in the former. Consequently, we vary the ionization parameter over a wide 
range to cover both classes of sources. We have produced
models with $\xi=1, 2, 5, 10, 20, 50,...,10^4,2\times 10^4, 5\times 10^4$~erg~cm~s$^{-1}$.
Finally, we also allowed the iron abundance to be treated as a free parameter, 
given the importance of the Fe K emission profile in most accreting sources.
For simplicity, the abundance of all the other elements considered, namely H, He, C,
N, O, Ne, Mg, Si, S, Ar, and Ca, is kept fixed to the solar values of \cite{gre98}.
Thus, models with $A_\mathrm{Fe}=0.5, 1, 5$ and $10$ were calculated, taking into account
both sub- and super-solar Fe abundances. Here, $A_\mathrm{Fe}=1$ corresponds to an iron abundance
of $2.5\times 10^{-5}$ with respect to hydrogen. These choices of parameter values resulted 
in a library of 720 synthetic 
reflection spectra that can be used for the modeling of the reprocessed component 
in X-ray observations. The present set of models is provided as a single file
\footnote{\url{http://hea-www.cfa.harvard.edu/~javier/xillver/}}
in FITS (Flexible Image Transport System) format, which can be loaded into the fitting 
package {\sc xspec}\footnote{\url{http://heasarc.gsfc.nasa.gov/xanadu/xspec/}}
via the {\tt atable} model.

\subsection{The effect of varying the ionization parameter $\xi$}

\begin{figure}
\epsscale{0.6}\plotone{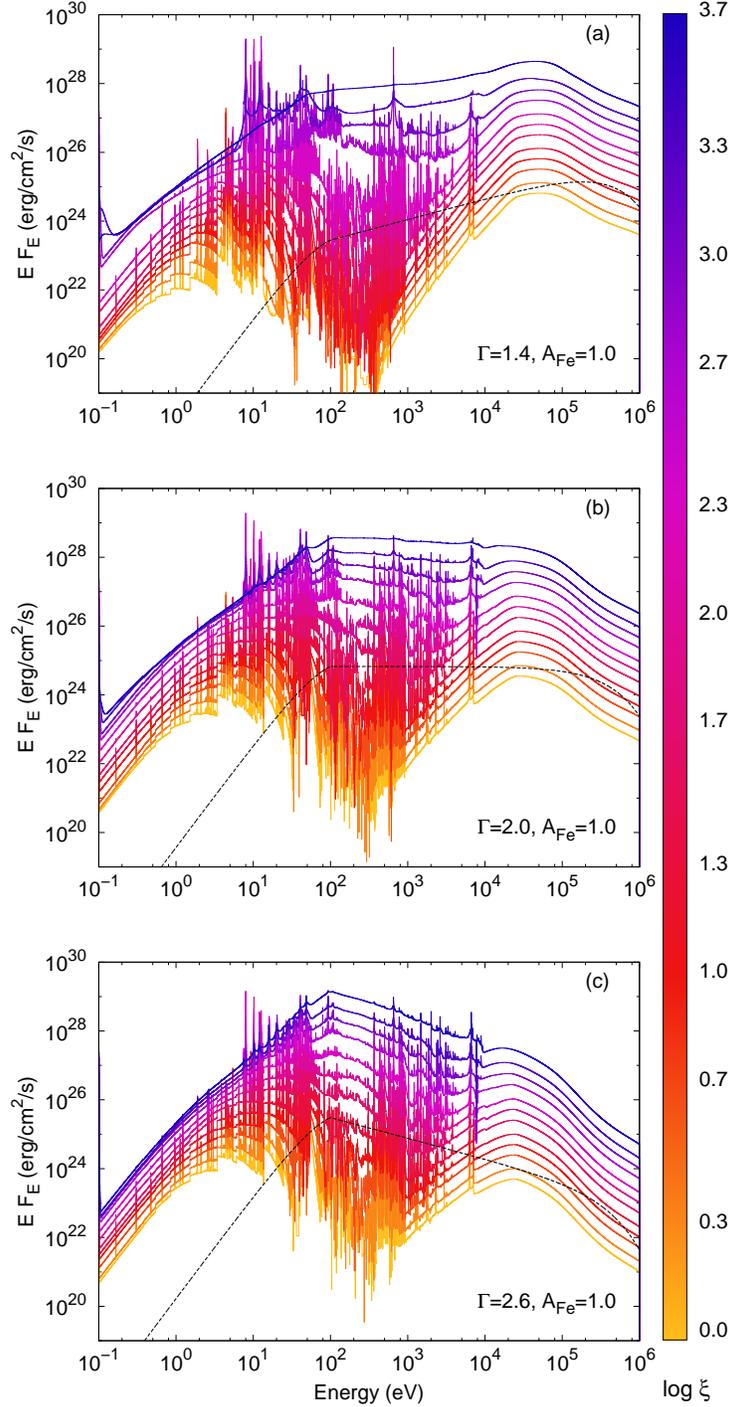}
\caption{Reflected spectra for different values of $\Gamma$ and $\xi$.
Panels (a), (b), and (c) correspond to $\Gamma=1.4$, $2.0$ and $2.6$, respectively. 
In each panel, solid lines show the reflected spectra calculated for a different value of 
the ionization parameter. From bottom to top, each curve corresponds to 
$\xi = 1, 2, 5, 10, 20, 50, 100, 200, 500, 1000, 2000, 5000$~erg~cm~s$^{-1}$, color-coded
accordingly. Note that no rescaling is applied. The illuminating power-law 
for the lowest value of $\xi$ is shown in black dashed lines.
The Fe abundance is set to the solar value ($A_\mathrm{Fe}=1$) for all the models.
}
\label{fspec.xi}
\end{figure}

Figure~\ref{fspec.xi} shows a sub-group of the resulting reflected spectra for three 
values of the photon index $\Gamma$ and a range of ionization parameters. The Fe abundance
is set to the solar value ($A_\mathrm{Fe}=1$) for all the models shown here. Panels (a), (b),
and (c) correspond to $\Gamma=1.4$, $2$ and $2.6$, respectively. Each panel shows
the reflected spectra calculated for a different value of the ionization parameter. From
bottom to top, each curve corresponds to $\xi = 1, 2, 5, 10, 20, 50, 100, 200, 500, 1000,
2000, 5000$~erg~cm~s$^{-1}$, respectively. Because the gas density is held fixed at $n_e=1.2\times 10^{15}$~cm$^{-3}$
for all these calculations, increasing the ionization parameter is equivalent to increase
the flux $F_x$ of the illuminating source, where $F_x = \xi n_e/4\pi$. The spectra are plotted
in units of $E F_E$ (equivalent to $\nu F_{\nu}$ if plotted in frequency), so that a power-law of
$\Gamma=2$ is shown as a horizontal line. Note that no rescaling or shift is applied;
instead, each curve is color-coded according to its corresponding value of log~$\xi$, 
to improve clarity. This Figure provides a general overview of the effect that the ionization
parameter has on the spectrum reflected from an optically thick plane-parallel slab.
The original power-law shape of the illuminating continuum (for the lowest value of $\xi$), 
shown in black dashed lines, suffers drastic modifications due to both absorption and 
emission. The spectra are shown in the entire energy range included in the calculations 
($0.1 - 10^6$~eV). In the low-energy part of the spectrum ($0.1-10$~eV), the continuum
is dominated by bremsstrahlung emissivity, despite the fact that this is where the 
illumination flux is decreased due to the cut-off imposed at $100$~eV. Emission lines
and absorption edges from H, He, and C are clearly visible for most of the models with
low and intermediate ionization. In the $10-10^4$~eV energy range, the photoelectric opacity
dominates over the electron scattering opacity. Therefore, this
region is where most of the absorption occurs, yielding large departures from the 
original power-law continuum. Bremsstrahlung emissivity decreases rapidly for energies 
above $\sim 100$~eV; however, many emission lines from all the ions included
in these calculations remain visible over the entire energy range. At higher energies ($>10^4$~eV),
electron scattering is the dominant source of opacity since the cross section for
photoelectric absorption decays as $\sim E^{-3}$, while the Klein-Nishina cross section 
for electron scattering remains fairly
constant. Thus, in this spectral region Compton scattering is the only relevant process and 
the reflected spectrum depends on the shape of the original illuminating field.

\begin{figure}
\epsscale{0.6}\plotone{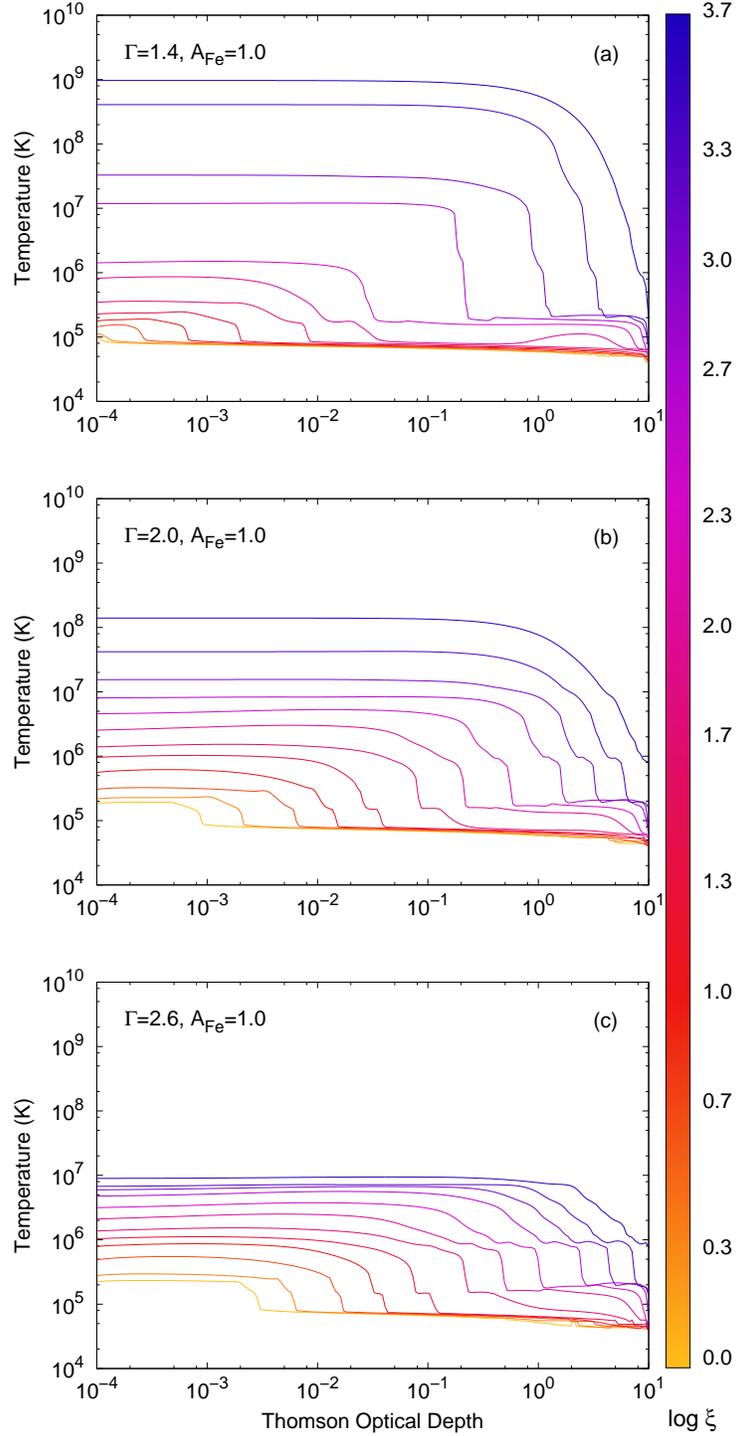}
\caption{Temperature profiles corresponding to the models shown in Figure~\ref{fspec.xi}.
Panels (a), (b), and (c) correspond to $\Gamma=1.4$, $2.0$ and $2.6$, respectively. 
In each panel, the solid lines represent the gas temperature along the vertical direction of the
slab, corresponding to a model with different value of the ionization parameter, color-coded
accordingly. The slab is illuminated at the surface ($\tau_{\mathrm T}=10^{-4}$), and the calculation
is carried out up to $\tau_{\mathrm T}=10$.
}
\label{ftemp.xi}
\end{figure}

By looking at the overall shape of the reflected continuum shown in Figure~\ref{fspec.xi}, 
it is possible to distinguish between two main regimes: (1) The {\it high-ionization case}, 
where the resulting spectra mostly shows very narrow emission features, while the continuum 
still resembles the original shape of the illuminating power-law; and (2) the 
{\it low-ionization case}, where the emerging spectra
are a combination of a very rich and complex set of emission-line profiles superimposed 
on a strongly absorbed and modified continuum, which significantly departs from the original
power-law. The specific value of the ionization parameter that separates these
two regimes depends, to a certain degree, on the photon index $\Gamma$. In fact, for the
$\Gamma=1.4$ case shown in panel (a) this transition is quite obvious, as a drastic
change in the reflected spectra can be seen between models with log~$\xi=2.3$ and $2.7$.
These changes are the result of large differences in the ionization balance solutions,
as can be seen in Figure~\ref{ftemp.xi}, which shows the corresponding temperature profiles 
for each one of the models shown in Figure~\ref{fspec.xi}, color-coded in the same way.
Because larger ionization implies a larger illumination flux, models with high $\xi$ values
are systematically hotter than those with low $\xi$ values. The illuminating radiation,
incident at the surface of the slab (at $\tau_{\mathrm T}=10^{-4}$ for our proposes), penetrates the
first layers heating the gas. The amount of heating not only depends on the illuminating
flux, but also on the particular shape of the radiation field; e.g., note that the models
for small values of $\Gamma$ are consistently hotter than those with large values.
We shall discuss these effects further in the next Section. As photons are
scattered or absorbed and re-emitted at different energies, the shape of the continuum is
modified reducing the net heating in the gas. At some point H and He recombine which increases
the cooling very rapidly. The gas then suffers a rather sudden transition to a lower temperature,
as can be seen for most of these models. This transition occurs at deeper regions in
the slab for larger ionization parameters and for softer input spectra (large $\Gamma$). 
Regardless of the ionization parameter or the shape of the ionizing
radiation, in all cases the field thermalizes and the gas eventually reaches the same
lower temperature around $(0.5-1)\times 10^5$~K. This is the limit in which the illuminating
radiation no longer contributes to the ionization of the material and its temperature
is simply set by the density of the gas.

\begin{figure}
\epsscale{1.0}\plotone{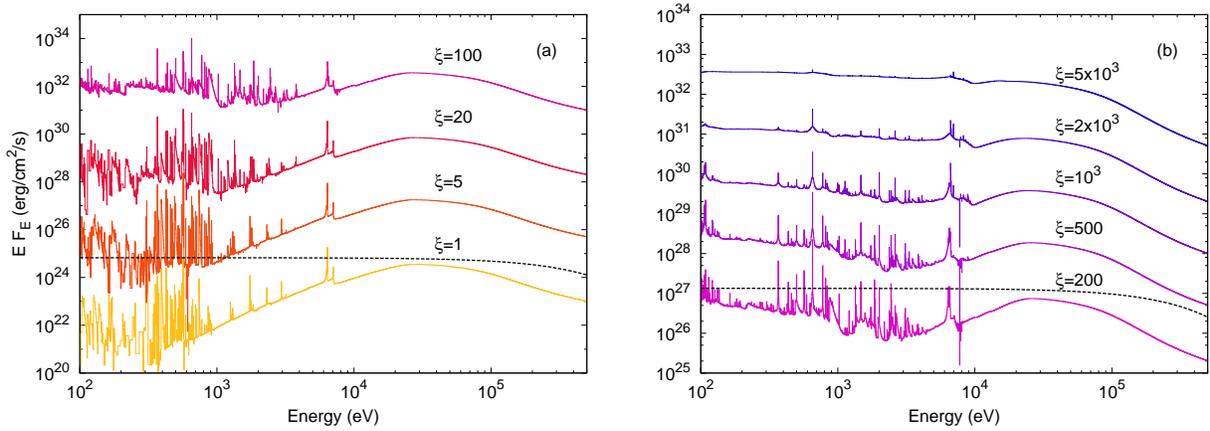}
\caption{Results showing the impact of the ionization parameter in the reflected spectra. 
In panel (a), solid lines show the models for $\xi=1, 5, 20,$ and $100$, multiplied by 
factors of $1, 10^2, 10^4,$
and $10^6$, respectively, to improve clarity. The dashed-line represents the incident power-law
for the model with $\xi=1$. In panel (b), solid lines are the reflected spectra for $\xi=200, 500,
1000, 2000,$ and $5000$, multiplied by factors of $1, 10, 10^2, 10^3,$ and $10^4$, respectively.
The dashed-line is the incident power-law for the $\xi=200$ case. In all cases, $\Gamma=2$
and $A_\mathrm{Fe}=1$.}
\label{flowhighxi}
\end{figure}

To illustrate in more detail the effects of the ionization parameter on the reflected spectra, 
we will concentrate on models with one particular value of the photon index.
Further, we will examine the low- and high-ionization models separately. Thus, 
Figure~\ref{flowhighxi} shows some of the reflected spectra resulting from models with 
$\Gamma=2$ in the $10^2 - 5\times10^5$~eV energy range, which is the spectral
band typically covered in X-ray observations. Panel~(a) shows 4 models with low-ionization,
specifically $\xi=1, 5, 20,$ and $100$, multiplied by factors of $1, 10^2, 10^4,$ and $10^6$,
respectively. Their corresponding temperature profiles are curves
1, 3, 5 and 7, from left to right, in panel~(b) of Figure~\ref{ftemp.xi}. The dashed line
represents the illuminating power-law spectrum for the model with the lowest ionization
parameter ($\xi=1$). As mentioned before, these models display a strong decrease of the
continuum flux due to photoelectric absorption, in particular for energies below $\sim 20$~keV.
Nevertheless, a very rich and complex set of fluorescent emission lines due to K-, L-,
and M-shell transitions from many ions is also present, superimposed on a highly 
absorbed continuum.  The general spectral shape of the lowest
ionization models ($\xi=1-20$) is very similar, where only small changes are seen. The
temperature in these models is relatively low ($T \lesssim 10^6$~K), and the gas settles
to a low temperature regime at a small optical depth ($\tau_{\mathrm T} \lesssim 0.1$), meaning
that most of the slab remains neutral. The emission due to the K$\alpha$ and K$\beta$ transitions
in Fe is distinctive at $\sim 6.4$ and $\sim 7.1$~keV, respectively. The case for $\xi=100$ 
is where the changes in the reflected spectra become more evident, in both the continuum 
absorption and in the emission features. The higher temperature in this model allows for the
excitation of elements such as Mg, Al, Si, S, Ar and Ca, which produces more fluorescence
lines in the $1-10$~keV region. The Fe K$\alpha$ emission becomes slightly broader and 
the K$\beta$ is less intense. Some radiative recombination continua (RRC) can also be 
seen near $\sim 0.5-1$~keV.

Panel~(b) of Figure~\ref{flowhighxi} shows the reflected spectra for the next five values
of the ionization parameter considered here ($\xi=200, 500, 1\times 10^3, 2\times 10^3,$ and
$5\times 10^3$). Each curve has been rescaled by a constant factor to improve 
clarity. These factors are $1, 10, 10^2, 10^3,$ and $10^5$ from low- to high-ionization, 
respectively. As before, the dashed line shows the illuminating power-law
spectrum corresponding to the model with the lower ionization ($\xi=200$). These are the
models considered as {\it high-ionization}, since the continuum is not highly affected by the
photoelectric opacity. This can be seen by comparing, for example, the reflected
flux at $10^2$~eV in the spectrum for $\xi=200$ and in the one for $\xi=1$, shown in panel~(a).
As mentioned before, increasing the ionization raises the temperature of the illuminated
region of the slab. This region also extends to deeper zones as the radiation is able to
ionize the gas at larger optical depths, as can be seen in the corresponding temperature
profiles (see  panel~b of Figure~\ref{ftemp.xi}). The increase in the temperature has two
main effects. On the one hand, it affects the ionization of the gas. The ions from low-$Z$ elements 
are completely stripped from all their electrons, while the ions of the heavier elements are partially ionized.
This changes the emission lines produced inside the gas, and thus the emerging spectra lacks
emission from low-$Z$ ions, progressively showing more emission from highly ionized O, Ne,
Ar, Ca and Fe, as the illumination increases. On the other hand, the high temperature affects
the way line photons are scattered by electrons, given the dependence of Equations~(\ref{eqec}) and
(\ref{eqsig}) on the kinetic energy of the electrons. When $4kT \sim E$, the probability for 
a photon to either gain or lose energy after each scattering becomes
comparable, which effectively produces a broadened and more symmetric line profile.
This effect is particularly evident in the K$\alpha$ emission lines from O and Fe, observed
at $\sim 0.65$ and $\sim 6.9$~keV, respectively. We shall return to a more detailed discussion
of the spectral features in Section~\ref{secfea}.

\subsection{The effect of varying the photon index $\Gamma$}

\begin{figure}
\epsscale{1.0}\plotone{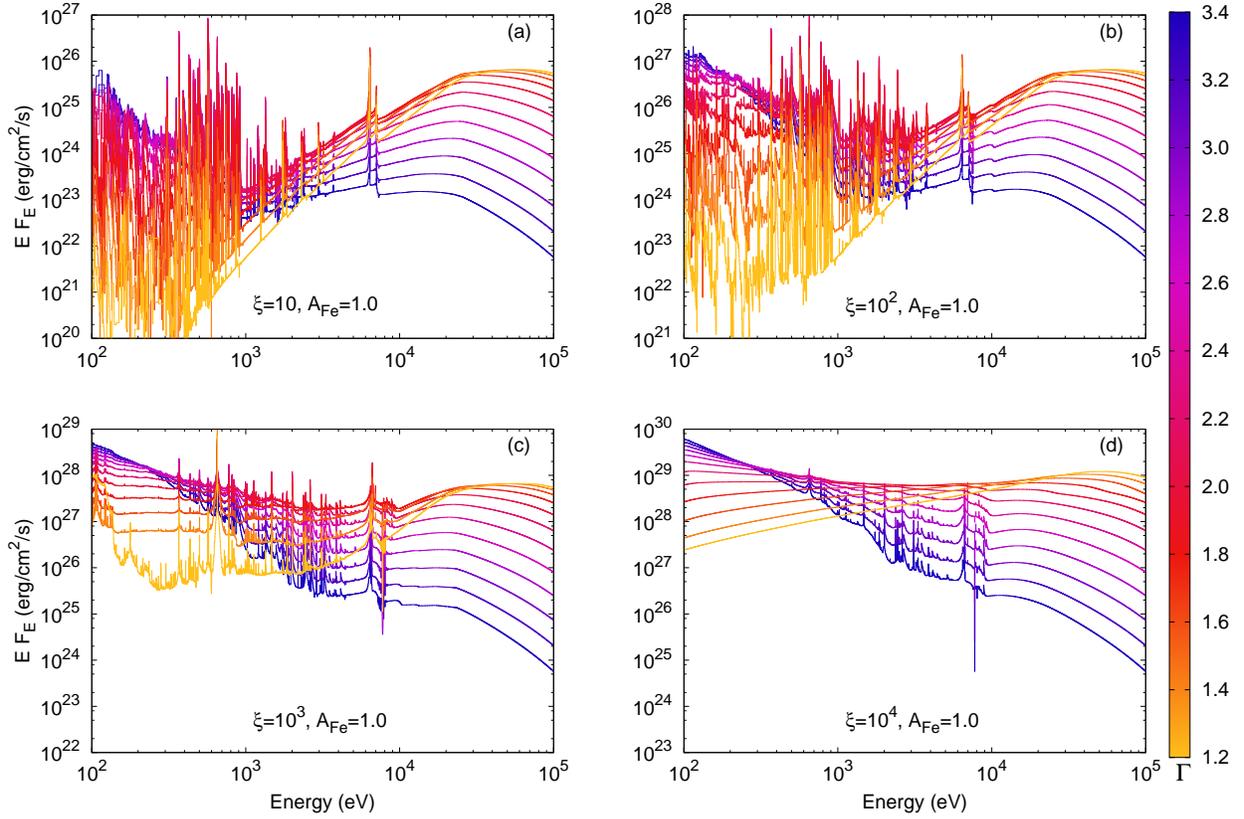}
\caption{Results showing the impact of the photon index in the reflected spectra.
Panels~(a), (b), (c) and (d) show the resulting models for a given ionization
parameter, i.e., $\xi=10, 10^2, 10^3$, and $10^4$, respectively. Each panel contains the
spectra of models with all the values of the photon index $\Gamma$ considered in our
calculations ($\Gamma=1.2 - 3.4$), color-coded accordingly. No rescaling is applied, and
$A_\mathrm{Fe}=1$ in all cases.
}
\label{fspec.gamma}
\end{figure}

It is clear from the discussion in the previous section that both the net flux and the 
spectral shape of the ionizing radiation incident on the surface of the slab have a great impact on the 
ionization balance of the gas, and thus on the reflected spectrum that will 
emerge at the surface. For this particular library of models we have adopted a
power-law shape for the illumination spectrum, thus the 
general shape is controlled by changing the value of the photon index $\Gamma$.
Figure~\ref{fspec.gamma} shows a sub-group of the reflected spectra calculated for various
conditions. Panels~(a), (b), (c) and (d) show the resulting models for a given ionization
parameter, i.e., $\xi=10, 10^2, 10^3$, and $10^4$, respectively. Each panel contains the
spectra of models for all the values of the photon index $\Gamma$ considered in our
calculations ($\Gamma=1.2 - 3.4$), color-coded accordingly. Since the goal of this Figure
is to show the general trends in the spectra introduced by changing $\Gamma$, all the curves
are plotted with the same normalization (thus, no rescaling was applied). It is quite obvious
how the changes in the ionizing continuum affect the resulting continuum
of the emergent radiation, as expected. However, the ionization structure of the gas
seems to follow different trends on $\Gamma$ for different values of $\xi$.

\begin{figure}
\epsscale{1.0}\plotone{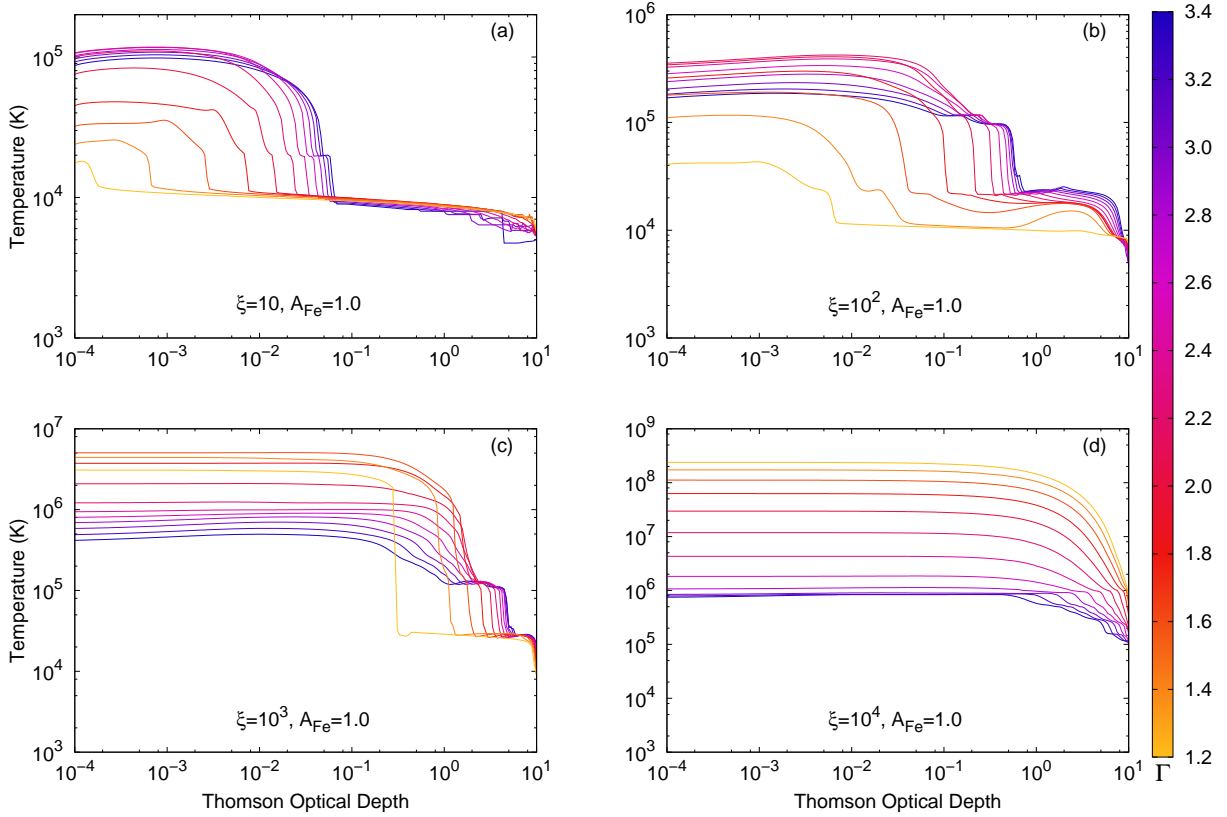}
\caption{Results showing the impact of the photon index in the temperature profiles.
Panels~(a), (b), (c) and (d) show the resulting models for a given ionization
parameter, i.e., $\xi=10, 10^2, 10^3$, and $10^4$, respectively. Each panel contains the
gas temperature along the slab for the models shown in Figure~\ref{fspec.gamma}, covering
all the values of the photon index $\Gamma$ considered in our
calculations ($\Gamma=1.2 - 3.4$), color-coded accordingly. $A_\mathrm{Fe}=1$ in all the
cases.
}
\label{ftemp.gamma}
\end{figure}

By comparing the two extreme values of $\xi$ (panels a and d), a completely opposite
behavior is seen. In the low ionization case ($\xi=10$), models with low $\Gamma$ values
are the ones with the most absorption, suggesting a much colder gas than those for high 
$\Gamma$ values. This is consistent with the temperature profiles obtained for these models, 
which are shown in panel~(a) of Figure~\ref{ftemp.gamma}. Therefore, in the low-ionization regime,
the softer the input spectrum (large $\Gamma$ value), the hotter and more ionized the
gas becomes. Looking at the high-ionization case in panel~(d) ($\xi=10^4$), the opposite
occurs: the models with harder illumination spectra (lowest $\Gamma$) are completely ionized,
owing to the large gas temperature. Thus, in the high-ionization regime, the harder the 
input spectra, the hotter and more ionized the gas becomes. The intermediate regimes plotted
in panels~(b) and (c) of Figures~\ref{fspec.gamma} and \ref{ftemp.gamma} show the transition
between these two regimes. For $\xi=10^2$ the gas temperature is low for the harder spectrum
($\Gamma=1.2$), and increases with the photon index up to $\Gamma \sim 2.4$, where it starts
decreasing again. For the case with $\xi=10^3$ (panel~c), the transition occurs at 
$\Gamma \sim 1.8$.

\begin{figure}
\epsscale{1.0}\plotone{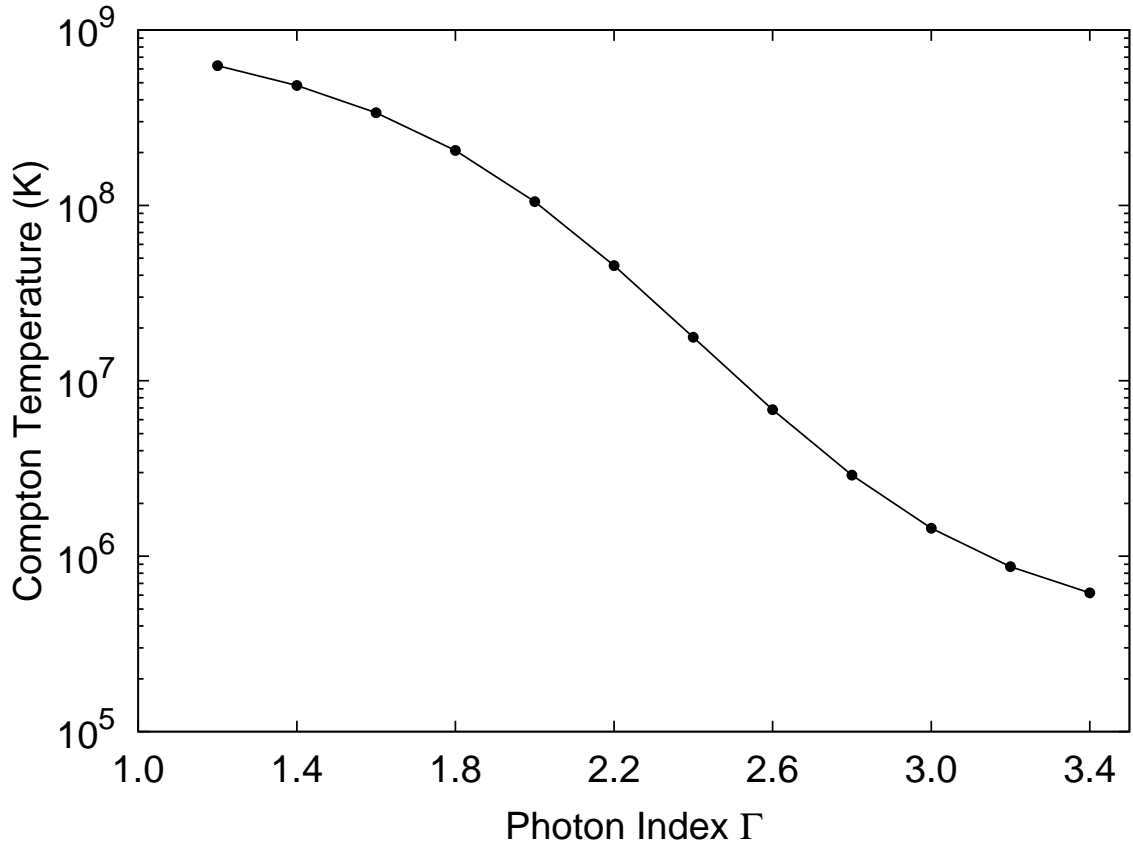}
\caption{Compton temperature as a function of the photon index $\Gamma$ for the illuminating
power-law spectra used in our calculations (see Equation~\ref{etcomp}).}
\label{ftcomp}
\end{figure}

The reason for this change of behavior is related to the processes that contribute to the heating
and cooling of the gas. In general terms, there are two main competing mechanisms: 
photoionization heating plus recombination cooling, and Compton heating and cooling
due to electron scattering. The low ionization models are those for which the illumination 
is relatively low. In this case, photoionization is the dominant process that heats the illuminated
layers of the slab. Because the heating rate due to photoionization is essentially given by
the radiation field flux times the photoelectric opacity of the gas, and since the latter
is dominant at energies below $\sim 10$~keV, a very hard input spectrum will produce much
less heating than a soft spectrum, owing to the lack of low energy photons. 
The contrary is true in the high-ionization regime where the dominant process is 
Compton heating and cooling. 
As discussed in Section~3.1 of \cite{gar10}, in this limit the gas temperature
approaches an asymptotic value, the Compton temperature, given by
\begin{equation}\label{etcomp}
T_C = \frac{<E>}{4k}
\end{equation}
where
\begin{equation}\label{emeane}
<E> = \frac{\int F(E)EdE}{\int F(E)dE}
\end{equation}
is the mean photon energy, which is a quantity that only depends on the spectral shape. 
In Figure~\ref{ftcomp} we show the resulting Compton temperature as a function of the
photon index $\Gamma$ in the range of our calculations, which agrees very well with the
temperature of the hot layer in the high-ionization models shown in panel~(d) of
Figure~\ref{ftemp.gamma}.\footnote{Note, however, that {\sc xstar} (and consequently {\sc xillver})
employs a full relativistic treatment of the Compton heating and cooling from \cite{gui82},
rather than using Equation~(\ref{etcomp}).} Physically this makes sense, since for the input spectra with 
low $\Gamma$ values most of the photons are concentrated in the high energy part of the 
spectrum, where Compton scattering becomes very important. Therefore, in the high-ionization
regime, hard spectra are more efficient in heating the illuminated layers of the slab.

\subsection{The effect of varying the Fe abundance}

\begin{figure}
\epsscale{0.8}\plotone{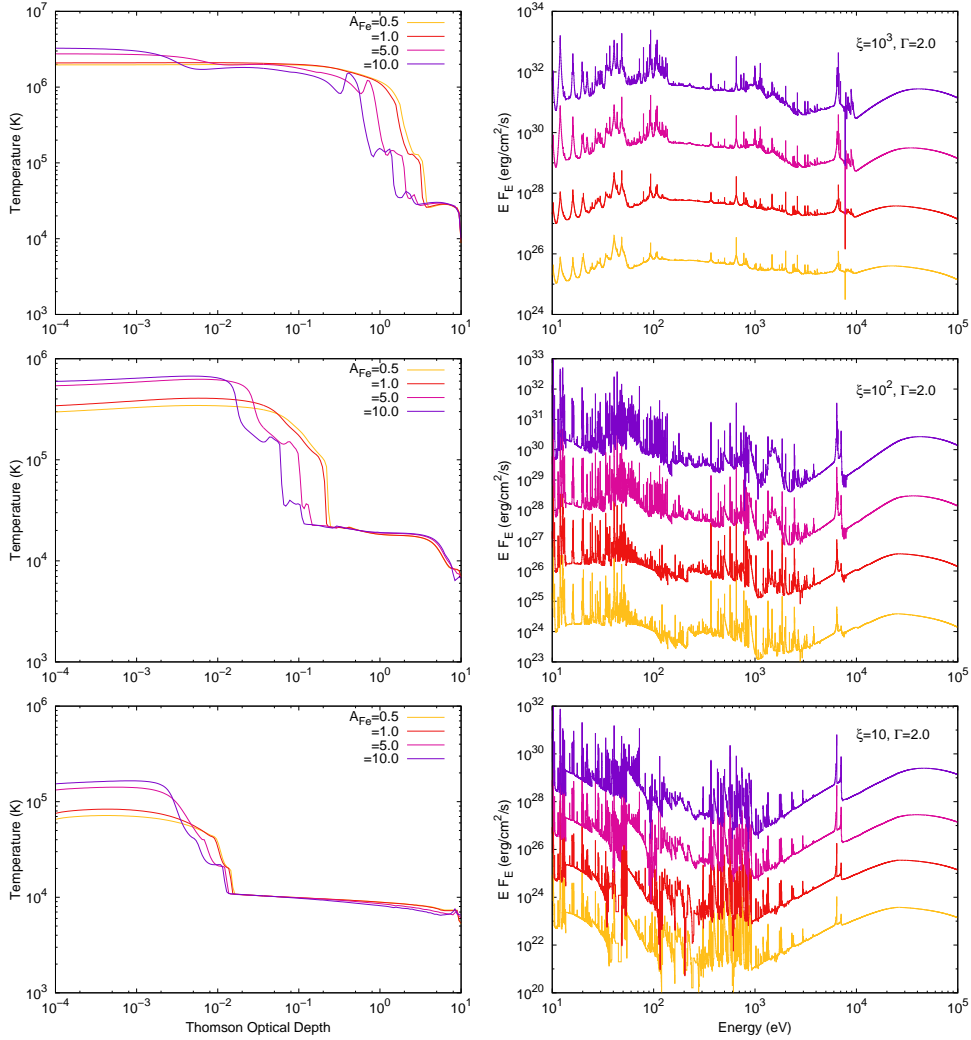}
\caption{Results showing the impact of the Fe abundance. Left panels show
the temperature profiles, and right panels the corresponding reflected spectrum.
Each curve corresponds to one particular value of $A_\mathrm{Fe} = 0.5, 1, 5$ and $10$. 
In each of the right panels, the plotted spectra have been rescaled for clarity. 
The scaling factors are, from bottom to top, $10^{-2}, 1, 10^2,$ and $10^4$. 
Top, medium, and bottom panels correspond to ionization parameters $\xi=10, 10^2$ and $10^3$, 
respectively. In all these models $\Gamma=2$.
}
\label{fafe}
\end{figure}

The elemental abundances considered in a photoionization calculation
can potentially affect the ionization balance, temperature structure, and ultimately
the observable spectral features in the reprocessed radiation. The total amount of
a particular element changes the continuum opacity, which in turn affects the photoionization
heating rate. At the same time, the abundance of a particular element 
influences the strength of the emission and absorption
features due to bound-bound and bound-free transitions. Given the relevance of the Fe
emission in the analysis of the X-ray spectra from accreting sources, we have carried
out calculations in which the Fe abundance, normalized to its solar value, is varied between
sub-solar, solar, and super-solar values. All the other elements considered in these calculations
are set to their solar values. Figure~\ref{fafe} shows a comparison of the reflection 
calculations for different values of the iron abundance $A_\mathrm{Fe}$. Left panels show
the temperature profiles, while right panels show the corresponding reflected spectrum
in the $10 - 10^5$~eV energy range. In all the panels, each curve corresponds to one 
particular value of $A_\mathrm{Fe} = 0.5, 1, 5$ and $10$, where $A_\mathrm{Fe}=1$ corresponds to
$2.5\times 10^{-5}$ of Fe with respect to H \citep{gre96}. In each one of the right panels, the 
plotted spectra have been rescaled for clarity. The scaling factors are, from bottom to top,
$10^{-2}, 1, 10^2,$ and $10^4$. Top, medium, and bottom
panels correspond to ionization parameters $\xi=10, 10^2$ and $10^3$, respectively.
The photon index is set to $\Gamma=2$ in all these models. The general tendency is the
same in all these simulations. The increase of the Fe abundance induces more heating
in the illuminated layers due to the increase in the opacity (and thus a larger photoionization
rate), which raises the gas temperature. However, because continuum absorption is also increased,
the radiation field thermalizes at a smaller depth for the high abundance models,
as can be seen from the temperature curves. Note that the increase in temperature 
is more subtle in the high-ionization case (top panel), since in this regime the Compton
heating and cooling is the dominant process that controls the gas temperature. 

The effects of the Fe abundance are evident in the reflected spectra as well,
both in the emission and the absorption features. The $10^2 - 10^4$~eV energy range clearly
shows a substantial reduction of the flux due to the increase in the continuum opacity;
meanwhile, the Fe K edge near $8$~keV grows deeper as $A_\mathrm{Fe}$ becomes larger. At the same time, 
all the Fe emission features are affected as well, which is mostly evident in the Fe K
emission complex in the $6-8$~keV region. The strength of the whole emission profile
increases when the abundance is high, as expected. In the high ionization models (top panel),
there is a distinctive RRC profile right before the Fe K edge, which becomes substantially
more prominent for $A_\mathrm{Fe}=10$. Also, there are enhanced emission lines near
$100$~eV and just above $1$~keV, which correspond to M- and L-shell transitions in iron,
respectively. However, these emission profiles are noticeable only for the $\xi=100$ model 
(middle panel), in particular the L-shell. The reason for this is that these transitions occur
in a rather narrow range of ionization stages. If the gas is very neutral, the photoelectric
absorption reduces considerably the number of ionizing photons at those energies, reducing
the number of excitations from the L-shell. On the contrary, if the gas is hot and ionized,
the fraction of Fe ions with L-shell electrons is very low (i.e., Fe~{\sc i}-{\sc xvi}),
as most of them are stripped already. This is of great relevance, since this feature
can be used to constrain both the ionization of the gas as well as the iron abundance.
One particular example is the Seyfert 1 galaxy 1H~0707-495, which X-ray spectrum shows
evidence of a very intense emission in both the Fe K- and L-shell regions. Fits using
reflection models require a high Fe abundance at an ionization parameter consistent with
the present analysis \citep{fab12b,dau12}.

Although somewhat extreme, 1H~0707-495 is not the only case where Fe is found to be over-abundant
based on predictions from reflection models. In fact, at least for AGN, 
it is commonly the case that super-solar iron abundance is required to fit the
observed X-ray spectra \citep{fab06}. A possible explanation for the apparent extreme Fe abundance
in some AGN has been proposed by \cite{rey12}, on the grounds of radiative levitation in the
accretion disk. If the radiative force exerted on a Fe ion exceeds the vertical gravity, this 
could cause iron to diffuse towards the photosphere of the disk, enhancing its abundance.   
But, in general, there is no particular reason why the 
other elements should be considered at their solar values, except the simplicity inherent 
in such an approximation. Given the capability of {\sc xillver} to treat any particular 
choice of elemental 
abundances, it can be used to produce smaller set of models custom made for any
specific situation. This could be of great use in peculiar systems such as ultra-compact
X-ray binaries, where a prominent O~{\sc vii} K$\alpha$ emission line observed in the X-ray
spectra is thought to originate from reflection in an accretion disk over-abundant in oxygen
\citep{mad10,mad11}. Also, emission lines from H-like S, Ar and Ca ions detected in some 
low-mass X-ray binaries suggest reprocessing material with compositions different from
solar \citep{dai09,dis09,egr12}.

\subsection{Spectral features}\label{secfea}

Figure~\ref{fspec.xi} shows the great complexity of the reprocessed
spectrum emerging from an illuminated, optically-thick
slab. Meanwhile, the models presented here constitute a
high-resolution representation of only one of the components observed
in the X-ray spectrum of accreting sources, namely the reflected
component.  In reality, one observes a composite spectrum that
includes the original power-law (presumably coronal) component plus
possibly a thermal blackbody-like component.  Additionally, if the
reflection occurs within a few gravitational radii from the compact
object, Doppler and gravitational redshifts will smear the spectral
profiles. Finally, absorption due to intervening gas such as warm
absorbers and outflows can also be present.  We shall discuss some of
the most prominent and representative of these features that one
expects to observe given the capabilities of current detectors.

\subsubsection{The Fe K-shell emission}

Undoubtedly, the emission complex near $6-7$~keV, which is due to
transitions from the inner K-shell of Fe ions, is the most prominent
atomic feature in the X-ray spectrum of accreting sources.  It is this
feature that provides the clearest evidence for reflection of
high-energy photons in the relatively cold, optically-thick material
of an accretion disk.  The ubiquity of this feature has been
established observationally for a large number of sources
\citep[e.g.,][]{got95,win09,ng10,fuk11} for the following two reasons.
First, the fluorescence yield, i.e. the probability of emission of a
photon rather than an Auger electron, is proportional to $Z^4$, where
$Z$ is the nuclear charge.  Second, the Fe K-shell lines are emitted
in a clean region of the X-ray spectrum ($6-8$~keV), where few other
ions emit or absorb radiation.  Furthermore, in this energy range
galactic absorption is negligible and most detectors operate quite
effectively.

\begin{figure}
\epsscale{0.8}\plotone{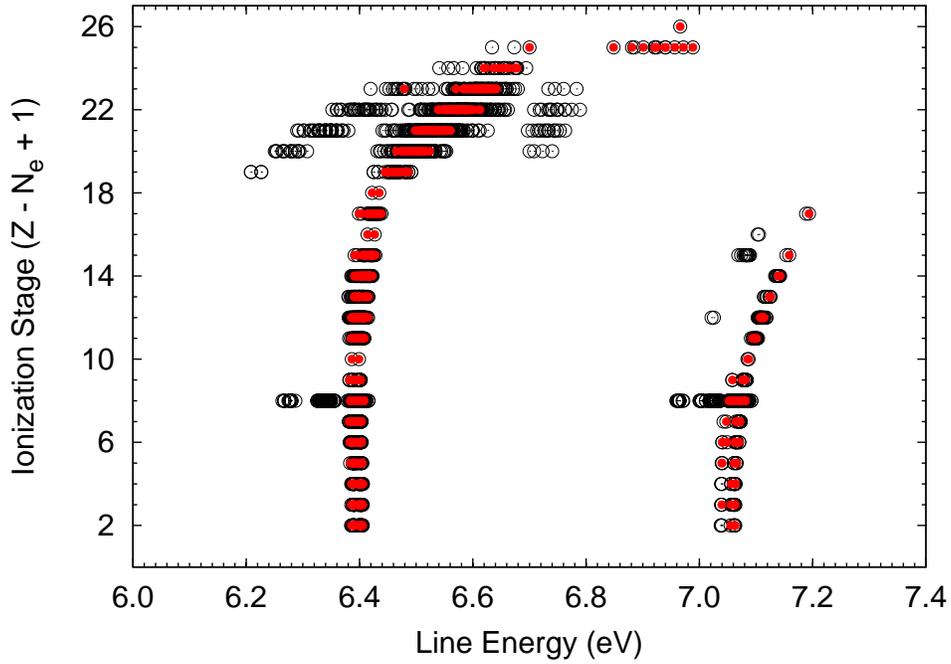}
\caption{Emission lines from all the Fe ions in the $6-10$~keV energy range (open circles). In the 
$x$-axis is the transition energy, and in the $y$-axis the ionization stage of each ion,
given by $Z-N_e+1$, where $Z$ is the nuclear charge and $N_e$ the number of electrons.
Filled circles show transitions with the highest probability ($A_r > 10^{13}$~s$^{-1}$).
Data accessible via http://heasarc.gsfc.nasa.gov/uadb/.
}
\label{falines}
\end{figure}

Figure~\ref{falines} shows a compilation of all the radiative
transitions from Fe ions in the $6-10$~keV energy range that are
included in our
database\footnote{\url{http://heasarc.gsfc.nasa.gov/uadb}}\citep[cf.,
Figure~3 in][]{kal04}.  There are a total of 2735 lines, all of which
correspond to K-shell transitions.  The open circles show the line
energy plotted against the ionization stage of each ion, as determined
by the relation $Z-N_e+1$, where $Z$ is the nuclear charge and $N_e$
is the number of electrons in the ion. Thus, an ionization stage of 2
corresponds to Fe~{\sc ii} (single ionized), 3 to Fe~{\sc iii} (double
ionized), and so forth. Filled circles show the most intense
transitions, i.e., those with large transition probabilities (here we
have chosen $A_r > 10^{13}$~s$^{-1}$).  Ionization increases upward:
The H- and He-like ions are near the top and the neutrals near the
bottom. The big group of points to the left corresponds to the
K$\alpha$ transitions ($n = 1 \rightarrow 2$, with $n$ being the
principal quantum number), and the smaller group to the right (for
which most energies are above $7$~keV) corresponds to the K$\beta$
transitions ($n = 1 \rightarrow 3$). Note that what is commonly
referred to as the neutral Fe K line at $6.4$~keV is in fact a
combination of several transitions that span the energy range
$6.39-6.43$~keV; this feature can be produced by many ions ranging
from Fe~{\sc i} up to Fe~{\sc xvii}. For Fe~{\sc xviii} and more
ionized ions, the transition energy spread is larger and the average
energy moves monotonically towards higher energies as the ionization
stage increases. For the He- and H-like ions (Fe~{\sc xxv-xxvi}),
there are only a few lines with energies around $6.9$~keV. Notice also
that the K$\beta$ transitions are only produced up to Fe~{\sc xvii},
since for higher ionization stages the $n=3$ shell is empty. The line
energy for the K$\beta$ transitions varies between $\sim 7.04$~keV for
Fe~{\sc ii} up to $\sim 7.19$~keV for Fe~{\sc xvii}.

\begin{figure}
\epsscale{0.8}\plotone{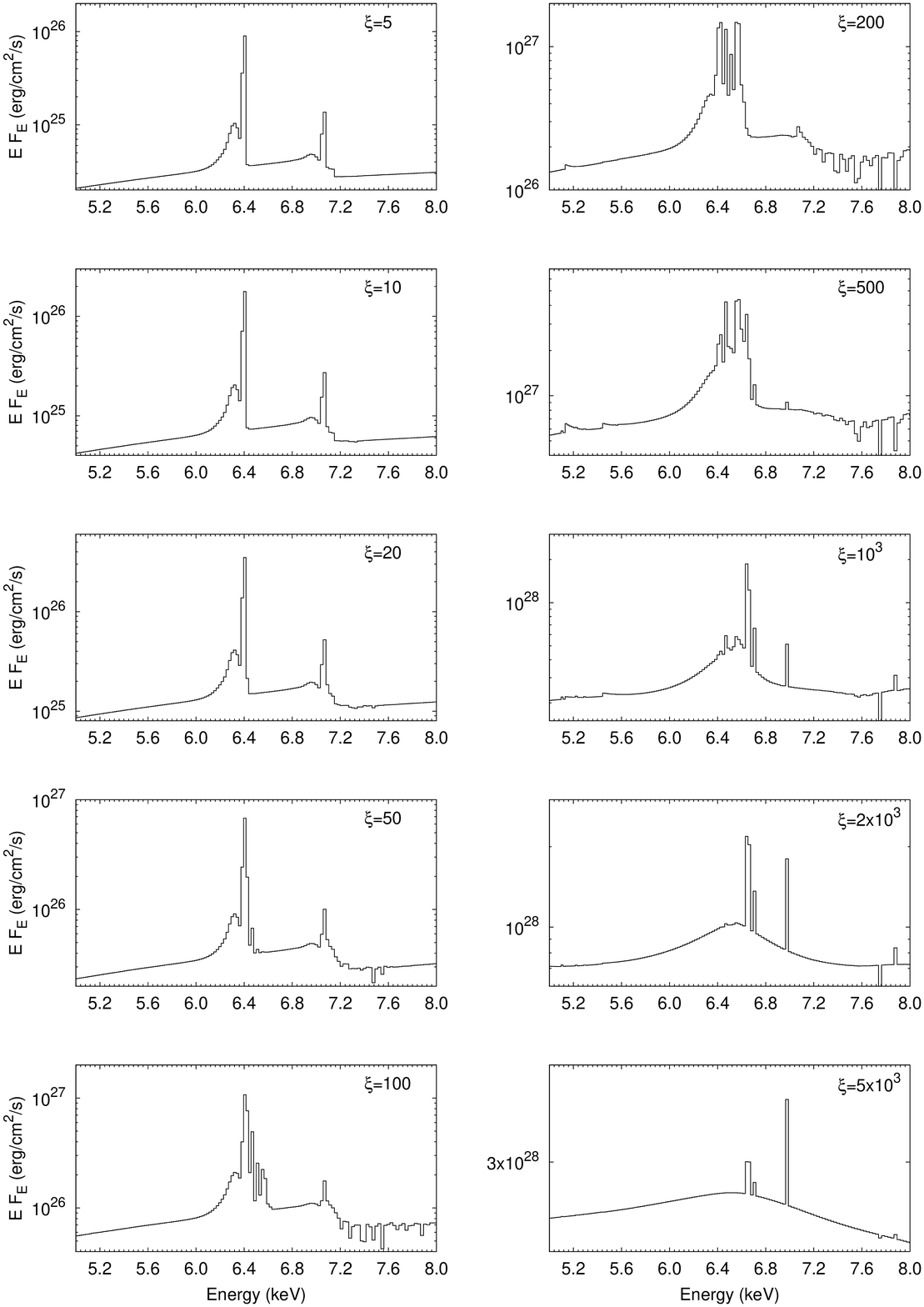}
\caption{Reflected spectra in the Fe K region ($5-8$~keV). Each panel corresponds to
a different ionization parameter, as indicated. In all these cases, $\Gamma=2$ and
$A_\mathrm{Fe}=1$.
}
\label{ffekspec}
\end{figure}

It is these line energies and intensities that we use to analyze the
emission in the Fe K region of the reflected spectra.
Figure~\ref{ffekspec} shows, in the $5 - 8$~keV band, the reflected
spectra for $\Gamma=2$ and $A_\mathrm{Fe}=1$ for different values of
the ionization parameter in the range $\xi=5 - 5\times 10^3$.
Ionization increases downward in the figure.  Starting at the lowest
level of ionization ($\xi=5-50$), the first few models show similar
groups of Fe lines, with two distinctive emission lines centered at
$\sim 6.4$~keV and $\sim 7.1$~keV, which correspond respectively to
the K$\alpha$ and K$\beta$ lines.  The line energies indicate that the
emission is dominated by Fe ions with ionization stages lower than
Fe~{\sc xvii}.  The smaller feature below $6.4$~keV, which is produced
by Compton down-scattering of line photons, is usually referred to as
the {\it Compton shoulder}.  For models with $\xi \gtrsim 200$, the
K$\beta$ line is no longer visible, which suggests that the emission
is dominated by ions at higher ionization stages.  In particular,
models with $\xi=200$ and $500$ show a rich complex of emission lines
at energies between $6.4-6.7$~keV.  At $\xi=10^3$, the emission is
centered at $\sim 6.7$~keV and $\sim 6.9$~keV, which implies that the
gas is highly ionized and most of the Fe ions are hydrogenic or
He-like.  In this state the temperature is high, and the overall line
emission profile is thermally broadened into a symmetric profile via
Compton scattering.  \cite{kal04} present a more detailed but similar
analysis of the Fe K emission complex based on {\sc xstar}
simulations, which includes atomic data for Fe; however, they do not
consider energy redistribution due to Compton scattering.

\subsubsection{Lower-$Z$ elements}
 
\begin{figure}
\epsscale{0.8}\plotone{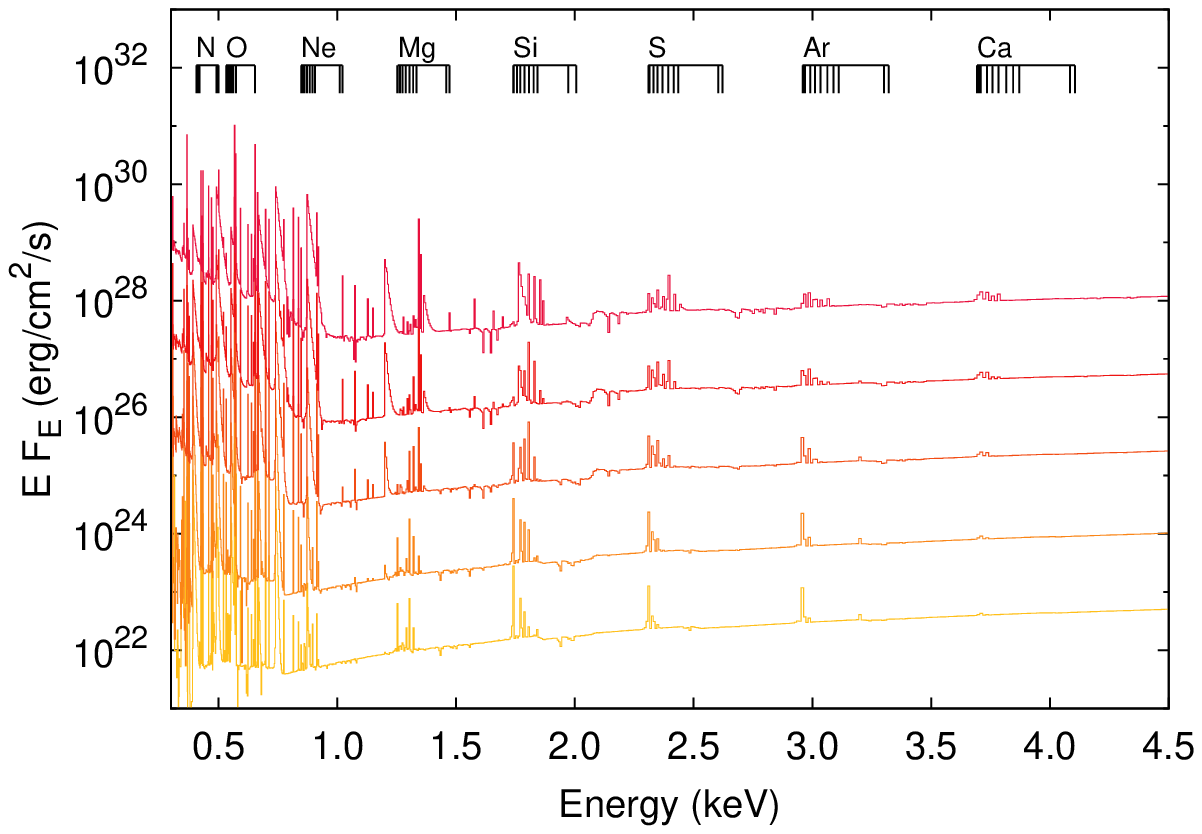}
\caption{Reflected spectra from models with steep illumination ($\Gamma=3$), and
solar abundances ($A_\mathrm{Fe}=1$). From bottom to top, each curve corresponds to 
$\xi=1, 2, 5, 10,$ and $20$, respectively. The strongest K$\alpha$ emission lines
are indicated for each ion considered in the calculation (except for neutral and single
ionized ions).
}
\label{flowz1}
\end{figure}

\begin{figure}
\epsscale{0.8}\plotone{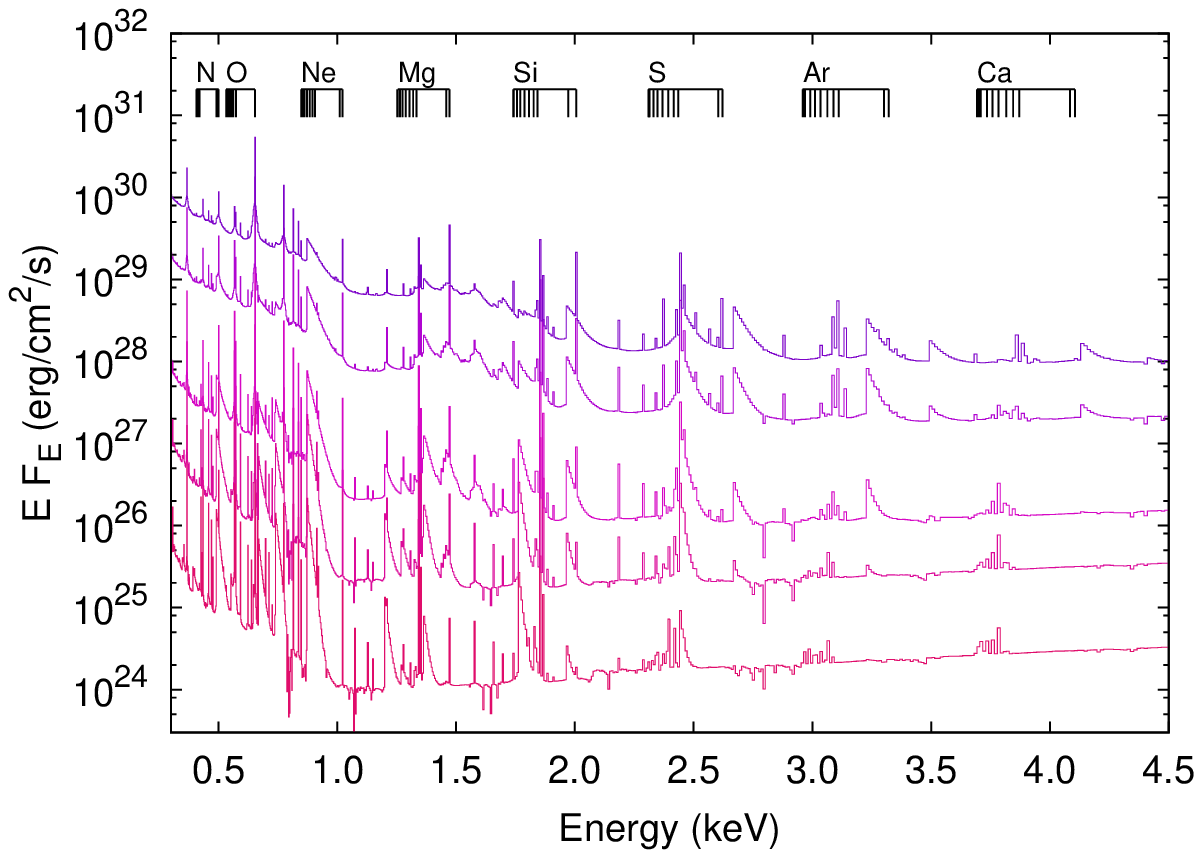}
\caption{Reflected spectra from models with steep illumination ($\Gamma=3$), and
solar abundances ($A_\mathrm{Fe}=1$). From bottom to top, each curve corresponds to 
$\xi=50, 100, 200, 500,$ and $10^3$, respectively. The strongest K$\alpha$ emission lines
are indicated for each ion considered in the calculation (except for neutral and single
ionized ions).
}
\label{flowz2}
\end{figure}

\begin{figure}
\epsscale{0.8}\plotone{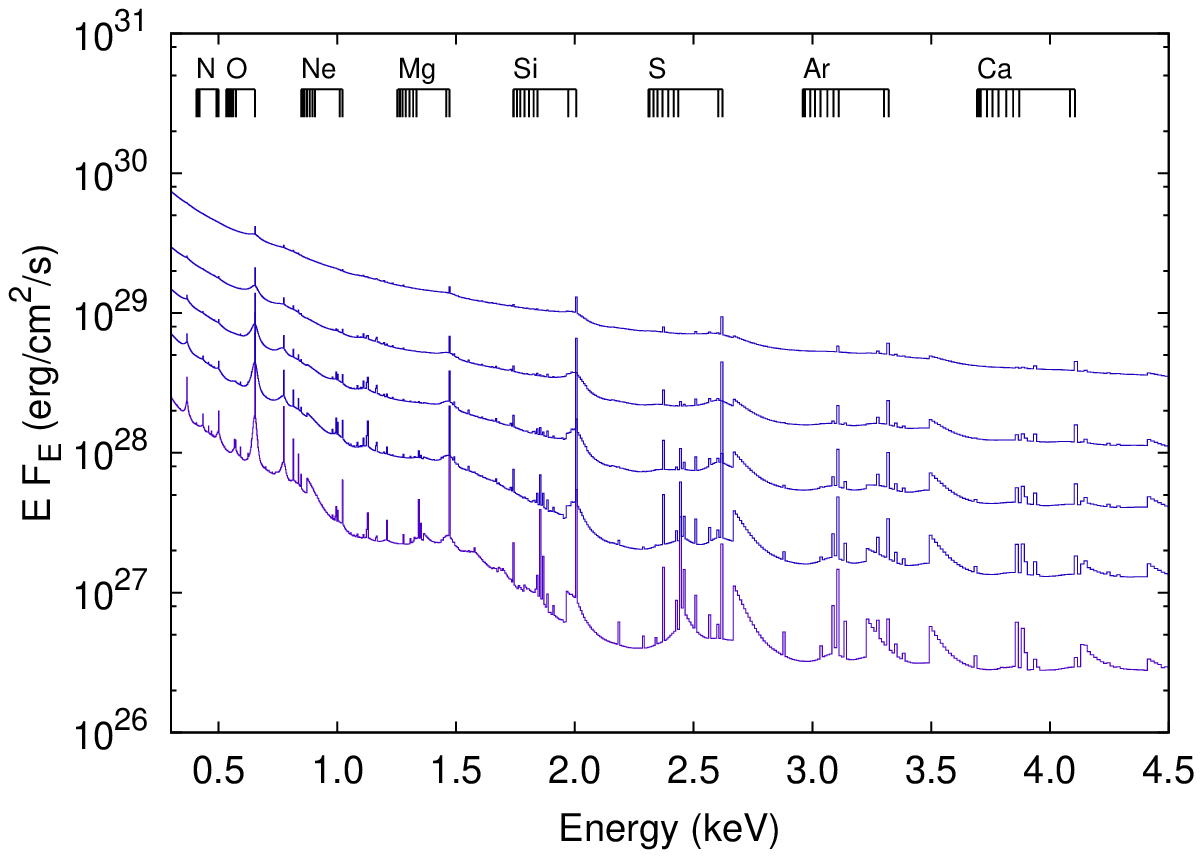}
\caption{Reflected spectra from models with steep illumination ($\Gamma=3$), and
solar abundances ($A_\mathrm{Fe}=1$). From bottom to top, each curve corresponds to 
$\xi=2\times10^3, 5\times10^3, 10^4, 2\times10^4$ and $5 \times10^4$, respectively. 
The strongest K$\alpha$ emission lines
are indicated for each ion considered in the calculation (except for neutral and single
ionized ions).
}
\label{flowz3}
\end{figure}

In addition to the important Fe-K emission complex in the reflected
spectrum, our code includes many emission lines of astrophysically
relevant ions at lower energies, which are due to such elements as C,
N, O, Ne, Mg, Si, S, Ar, and Ca.  The intensities of these lines tend
to increase as the illuminating spectrum softens (i.e., as $\Gamma$
increases) and as the flux correspondingly increases at lower energies
where photoelectric absorption dominates.  Figures~\ref{flowz1},
\ref{flowz2}, and \ref{flowz3} show the reflected spectra for
$\Gamma=3$ and $A_\mathrm{Fe}=1$ and for all the values of the
ionization parameter considered in our library.  The spectra cover the
$0.3 - 4.5$~keV energy range, which is the band that contains most of
the inner-shell transitions from low-$Z$ ions.  At the top of each
figure is indicated the energy of the strongest K$\alpha$ emission
line for each ion in the isonuclear sequence of each of the elements
contained in our atomic database (except for the neutral and
single-ionized cases).  While these K$\alpha$ lines are not at all the
only lines considered in our models, they nevertheless give an
indication of the energy range one expects most of the features of a
given element to appear.

Figure~\ref{flowz1} shows spectra for the lowest ionization models,
namely $\xi=1, 2, 5, 10,$ and $20$ (with ionization increasing from
bottom to top).  At the lowest energies ($0.3 - 1$~keV), the lines are
indistinguishably blended together.  At higher energies, emission
lines from low ionization states of Mg, Si and S are evident, and
relatively weak lines due to Ar and Ca K$\alpha$ are also present.
The radiative recombination continuum (RRC) due to H-like Ne at $\sim
1.4$~keV becomes visible in models with $\xi \gtrsim 5$.
Such features are produced by recombining electrons with
energies that exceed the ion binding energy.  The excess energy is
radiated as a photon.  The energy of a typical RRC photon is $E =
E_{IP} + kT_e$, where $E_{IP}$ is the ionization potential of the ion
and $T_e$ is the electron temperature.

For higher ionization parameters, the reflected spectra are dominated
by RRC features, as shown in Figure~\ref{flowz2}, which displays
spectra ranging from $\xi=50$ at the bottom to $\xi=10^3$ at the top.
Also visible are higher-excitation emission lines of Mg, Si, S and Ar.
The H-like oxygen RRC at $\sim 0.87$~keV is quite strong for $\xi=50$
(bottom curve), and it weakens as $\xi$ increases.  Strong emission
features due to K$\alpha$ transitions of Mg, Si, and S appear at $\sim
1.4, 1.8,$ and $2.45$~keV, respectively.  The latter feature
corresponds to a blend of the S~{\sc xv} K$\alpha$ line and the
Si~{\sc xiii} RRC \citep{gar11}.

Figure~\ref{flowz3} shows the last five models of the series with the
ionization parameter increasing from $\xi=2\times 10^3$ (bottom) to
$\xi=5\times 10^4$ (top).  For these large values of $\xi$, fewer
emission lines are present.  However, they are quite intense because
the soft illuminating spectrum is rich in low-energy photons.  The
strongest emission lines are produced by H-like ions of O, Mg, Si, and
S with energies of $\sim 0.654, 1.47, 2.01,$ and $2.63$~keV,
respectively.  The RRC from H-like Si and S are clearly seen at $\sim
2.67$ and $3.5$~keV, respectively.  Generally, K$\alpha$ fluorescence
from the low ionization states of O, Mg, Si, S, Ar, and Ca ions is
most important at low column depths because of the higher K shell
opacity of these elements relative to Fe.  Conversely, Fe K$\alpha$ is
more important at relatively high columns, i.e., $\tau_{\mathrm T}\sim
1$.

\subsection{Comparison with previous models}

Considering the several reflection models currently available, {\sc
reflionx} \citep{ros05} is the one that is most similar to the models
presented in this paper.  It has been widely used by the scientific
community in analyzing spectral data for many observations of various
black-hole and neutron-star sources.  We therefore benchmark our
results by making a detailed comparison of {\sc xillver} and {\sc
reflionx}, while describing the advantages of our model.

In {\sc reflionx}, the radiation field is separated into the direct
component of the illuminating radiation, which is unaffected by either
scattering or absorption (assumed to be $\propto e^{-\tau_{\nu}}$),
and the diffuse component, which results from Compton scattering and
emission within the slab. The Compton processes are treated by solving
the Fokker-Planck diffusion equation, which includes a modified
Kompaneets operator\citep{ros78,ros79}.  Incident photons that are
Compton scattered in the slab contribute to the emissivity of the
diffuse field, and it is assumed that their distribution in energy is
described by a Gaussian with central energy and energy dispersion
given by Equations~(\ref{eqec}) and (\ref{eqsig}).  Due to the
deficiency of the Fokker-Planck equation in handling steep gradients
in the energy spectrum, resonance lines are treated using the escape
probabilities technique.  In addition to fully ionized species, the
\cite{ros05} calculations include: C~{\sc iii-iv}, N~{\sc iii-vii},
O~{\sc iii-viii}, Ne~{\sc iii-x}, Mg~{\sc iii-xii}, Si~{\sc iv-xiv},
S~{\sc iv-xvi}, and Fe~{\sc iv-xxvi}.  However, none of the neutral or
near-neutral ions are included in their models.  The cross sections
for photoionization are calculated from the fits of \cite{ver95}; in
the case of Fe, transition probabilities and Auger decay rates are
taken from \cite{kas93}.  The items we have just highlighted are among
the most important differences between the atomic data employed in the
two models (see Section~\ref{secatomicd}).  It is also important to
mention that the Fe K$\alpha$ lines treated in {\sc reflionx} are the
recombination lines of Fe~{\sc xxvi} and Fe~{\sc xxv}, and the
fluorescence lines of Fe~{\sc vi-xvi}, while the K$\alpha$
fluorescence of Fe~{\sc xvii-xxii} is assumed to be suppressed by
autoionization. All K$\beta$ and higher $n$ resonances are also
neglected.

\begin{figure}
\epsscale{1.0}\plotone{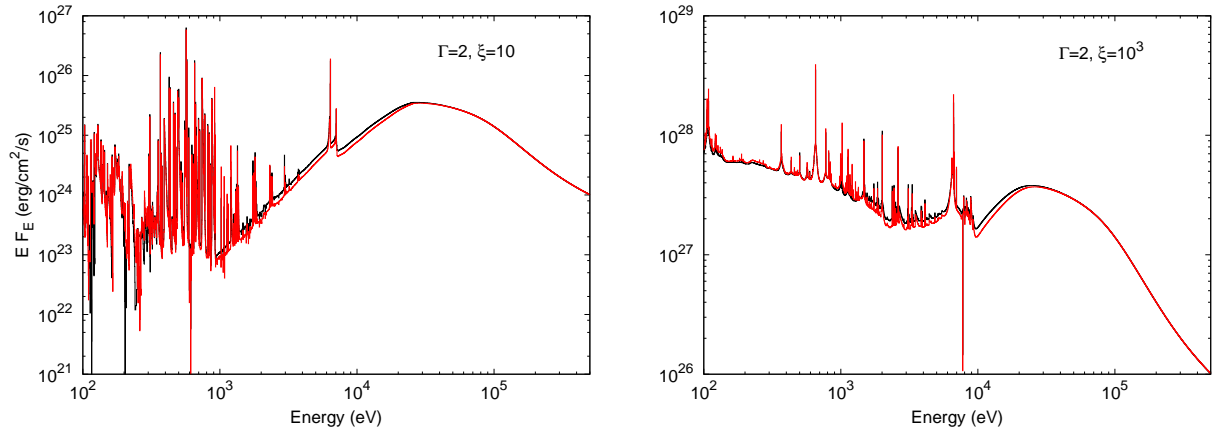}
\caption{Reflected spectra from {\sc xillver} using the elemental abundances of \cite{gre98}
(black curve), and those by \cite{mor83} (red curve). Left panel shows the case for $\xi=10$,
while right panel shows the $\xi=10^3$ models. In all cases $\Gamma=2$.}
\label{fabund}
\end{figure}

The solar abundances of the elements in {\sc reflionx} are taken from
\cite{mor83}, while in {\sc xstar}, and thus in {\sc xillver}, we
adopt the values of \cite{gre98}.  Table~\ref{tabund} shows the values
reported from both sources for each element, and the last column shows
the ratio of the two values.  Apart from N and Ar, all the abundances
used in {\sc xillver} are lower than those used in {\sc reflionx}.  In
particular, O is lower by $\sim 10\%$ and Fe by $\sim 30\%$, while,
most notably, Ne is $\sim 80\%$ lower.  The choice of abundance model
is relatively unimportant, as illustrated in Figure~\ref{fabund},
which shows reflected spectra computed using {\sc xillver} for
$\xi=10$ (left panel) and $\xi=10^3$ (right panel).  Results obtained
using the model of \cite{gre98} are shown in black, and those obtained
using the model of \cite{mor83} in red.  The differences in each panel
are small.  For the higher \cite{mor83} abundances, the continuum is
slightly depressed and some of the strongest lines are somewhat
weaker.  These small differences are unlikely to be important in
analyzing real data.

\begin{figure}
\epsscale{0.6}\plotone{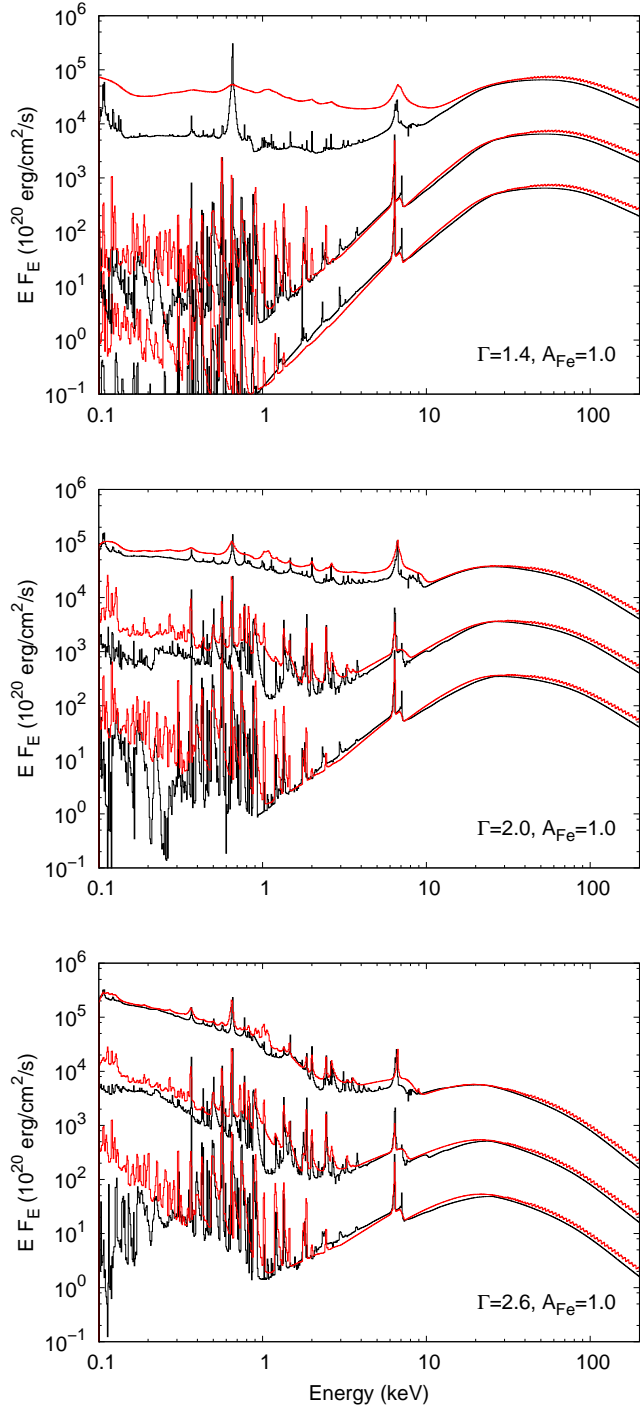}
\caption{Comparison of the reflected spectra calculated with {\sc xillver} (black curves),
and with {\sc reflionx} (red curves). Top, middle, and bottom panels show the models for
$\Gamma=1.4, 2,$ and $2.6$, respectively. In each panel, pairs of curves correspond to
$\xi=10, 10^2,$ and $10^3$, from bottom to top. All spectra are plotted with the same 
energy resolution. Note that $A_\mathrm{Fe}=1.32$ in {\sc xillver} to compensate for the difference
in the solar values with those used in {\sc reflionx}.}
\label{fcompref}
\end{figure}

Although the solar abundances assumed for both {\sc reflionx} and {\sc
xillver} are fixed and different, both models allow the Fe abundance
to be varied.  This allows us to compensate for the abundance
differences in making direct comparisons of the two models.  One such
set of comparisons is shown in Figure~\ref{fcompref}, where the
emergent spectra for different values of both the ionization parameter
and the photon index are plotted for {\sc xillver} (in black), and
{\sc reflionx} (in red).  Top, middle and bottom panels are for
$\Gamma=1.4$, $2$, and $2.6$, respectively.  Each panel shows 3 pairs
of curves, corresponding to $\xi=10, 10^2,$ and $10^3$, from bottom to
top, respectively.  The spectra are from the FITS tables that are
accessible via the {\tt atable} model in XSPEC.  To ensure that the
energy resolution is the same for all the spectra, in all cases we
used a logarithmic dummy response of 1000 energy points over the
energy range $0.1 - 1000$~keV.  Furthermore, appropriate grid point
were chosen in the parameter space to avoid problems that might be
introduced by the interpolation procedure.  The normalization for both
models is the same because we use consistent definitions for the
ionization parameter and the power-law spectrum.  Thus, all the
parameters were set to be identical for the two models, with the
exception of the iron abundance: It was set to $A_\mathrm{Fe}=1.32$ in
{\sc xillver} in order to compensate for the difference in the solar
values assumed in each model, as discussed above.

In general, the two models are in better agreement for low ionization
and soft spectra, as in the models for $\Gamma=2.6$ shown in the
bottom panel of Figure~\ref{fcompref}.  However, even in these cases
there are important differences: For the bottom spectra with $\xi=10$,
the {\sc xillver} model is more absorbed at low energies than the {\sc
reflionx} model.  While the energies and intensities of the strongest
lines are in good agreement, some weaker features at $\sim 2$~keV and
$\sim 3$~keV are only present in the {\sc xillver} spectrum, probably
due to the differences in the atomic data.  These particular features
are much stronger for the harder spectra shown in the middle and top
panels.  The differences in the atomic data also affect the Fe K
emission profile near $6-8$~keV.  For example, Fe K$\alpha$ emission
at $\sim 6.4$~keV is somewhat more intense in our model and K$\beta$
is conspicuously absent in the {\sc reflionx} spectra.  Concerning the
Fe continuum, one might expect significant differences because the
models use very different photoionization cross sections.  However,
the models agree very well on the depth of the Fe K edge near
$7.2$~keV, especially at low ionization.  As mentioned above, the
discrepancies are most pronounced for high ionization and for the hard
spectra shown in the top panel ($\Gamma=1.4$).  The difference for the
pair of curves for $\xi=10^3$ is remarkable: The {\sc xillver}
continuum below $\sim1$ keV is at least an order of magnitude fainter
than the {\sc reflionx} continuum; the extremely intense O Ly$\alpha$
emission line at $\sim 0.65$~keV in our spectrum is scarcely present
in the {\sc reflionx} spectrum; and the Fe K line in the latter
spectrum is much broader than in the former.

\begin{figure}
\epsscale{0.8}\plotone{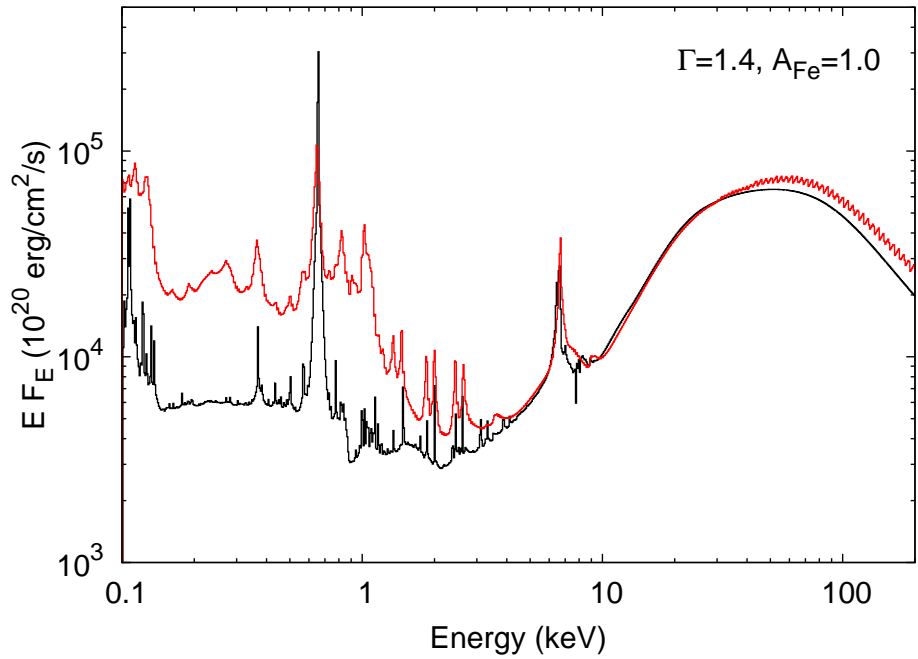}
\caption{Reflected spectra for $\Gamma=1.4$ and solar abundances. In black is the
{\sc xillver} result for $\xi=10^3$. In red is the {\sc reflionx} spectra for $\xi=500$
and twice the normalization. Although the Fe K emission agrees better, there are large
discrepancies in the reflected flux at lower energies.}
\label{fcompref2}
\end{figure}

These large discrepancies for high ionization and hard spectra are
unexpected and there is no ready explanation.  The lack of emission
lines from low-$Z$ elements suggests that (for the same input
parameters) the slab is hotter and more ionized for {\sc reflionx}
than for {\sc xillver}.  We test this idea by comparing the
performance of the two models for different values of the ionization
parameter: In Figure~\ref{fcompref2} we compare the two $\Gamma=1.4$
models where for {\sc reflionx} we have doubled the normalization and
reduced the ionization parameter to $\xi=500$ (half the value for {\sc
xillver}), in order to match the fluxes at the higher energies.  The
shape and intensity of the Fe K line is now in better agreement.
However, the {\sc reflionx} continuum is still considerably higher at 
low energies; as a consequence, the reflected spectrum is even softer than the
incident power-law spectrum.

\begin{figure}
\epsscale{0.8}\plotone{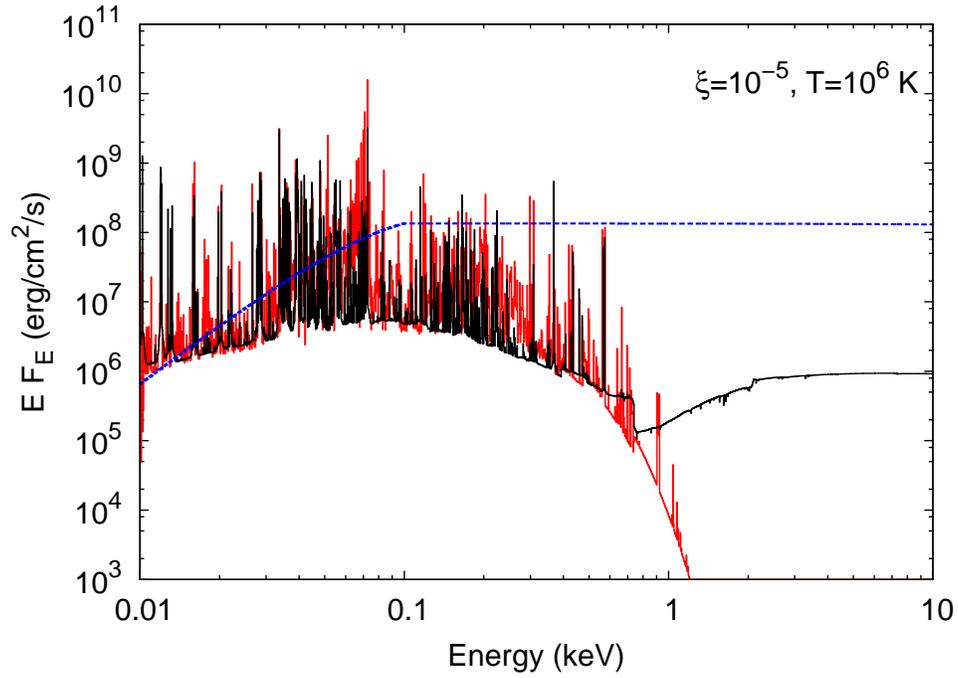}
\caption{Comparison of the emergent spectra from a collisional plus bremsstrahlung
dominated gas. In red is the spectrum from the {\sc apec} model with $T=10^6$~K and
solar abundances. In black is the {\sc xillver} prediction for a thin slab ($\tau_{\mathrm T}=10^{-2}$),
at the same temperature, and log~$\xi=-5$. The blue dashed line shows the incident spectrum.}
\label{fapec}
\end{figure}

In order to test whether our code is reliably modeling this low-energy
continuum, we compared our results with those computed using the {\sc
apec} model \citep{fos12}.  For this comparison, with {\sc xillver}
we computed the reflection spectrum from a thin slab ($\tau_{\mathrm
T}=10^{-2}$) for solar abundances at very low ionization with the
temperature held fixed at $10^6$~K.  These parameters closely match
the case of a collisionally-ionized gas whose spectrum is dominated by
bremsstrahlung emission.  Figure~\ref{fapec} shows the comparison.
Setting aside the differences in the emission lines (a result of the
differences in the atomic data sets), the spectra are in good
agreement, especially the level of the continuum.  This comparison
gives us confidence in {\sc xillver}'s implementation of the free-free
emissivity.  

\begin{figure}
\epsscale{0.6}\plotone{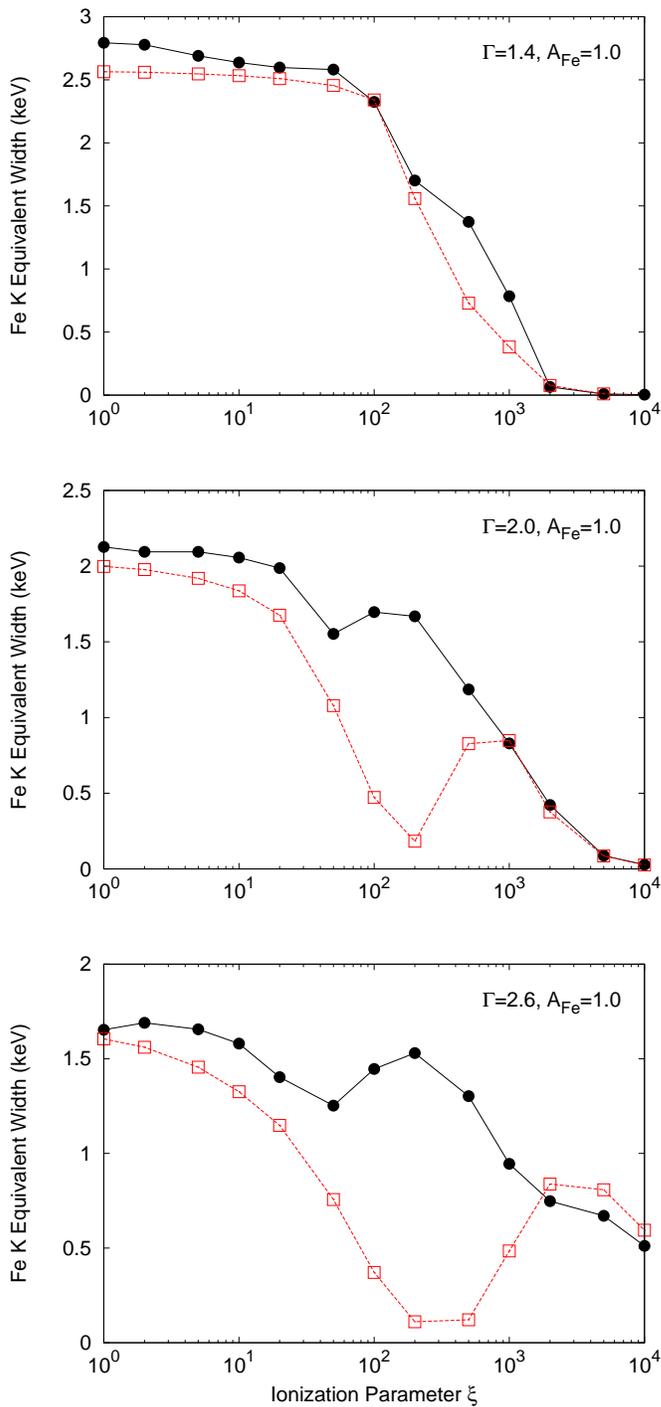}
\caption{Fe K emission line equivalent width versus ionization parameter obtained from
the spectra calculated by {\sc xillver} (filled circles), and {\sc reflionx} (open squares).
Top, middle, and bottom panels correspond to models with $\Gamma=1.4, 2,$ and $2.6$, respectively.
}
\label{fews}
\end{figure}

An important quantity to compare these models is the equivalent
width (EW) of the Fe K emission complex.  Following \cite{gar11}, we
use the well-known formula,
\begin{equation}\label{eew}
EW= \int_{5.5~\mathrm{keV}}^{7.2~\mathrm{keV}}\frac{F(E) - F_c(E)}{F_c(E)}dE,
\end{equation}
where $F(E)$ is the flux of the reflected spectrum and $F_c(E)$ is the
flux in the continuum.  The $5.5 - 7.2$~keV range of integration
covers all Fe emission features, including the K$\beta$ lines.  The
continuum is approximated by a straight line that passes through the
endpoints of the energy band.  Clearly, this straight-line continuum
and this particular choice of integration limits are somewhat
arbitrary and by no means constitutes an accurate determination of a
physical quantity, but it suffices to compare the two models.
Figure~\ref{fews} shows plots of the EWs vs. the ionization parameter
for both {\sc xillver} and {\sc reflionx}, with $\Gamma$ increasing
from top to bottom.  For both models, the EWs tend to decrease as the
ionization increases.  At low ionization, the EWs are similar,
although the values for {\sc xillver} are consistently higher, which
is expected because the {\sc reflionx} model lacks the Fe K$\beta$
lines.  Meanwhile, at high ionization ($\xi \gtrsim 10^3$), where the
Fe K emission is dominated by the H- and He-like ions, the agreement
in the EWs is also good, but values for the {\sc reflionx} model are
in this case somewhat greater for large $\Gamma$ (bottom panel).

Significant discrepancies appear when one compares the two models at
higher levels of ionization, $10^2 \lesssim \xi \lesssim 10^3$: The
EWs for the {\sc reflionx} models decrease drastically, especially for
the softer spectra (bottom panel).  This behavior may occur because
{\sc reflionx} does not include the Fe K$\alpha$ lines for most of the
second-row ions (Fe~{\sc xvii-xxii}) under the assumption that these
lines are suppressed by Auger resonant destruction \citep{ros96}.
{\sc xillver}, on the other hand, predicts strong K$\alpha$ emission
at the energies at which many of these ions are expected to emit
(e.g., see Figures~\ref{falines} and \ref{ffekspec}).  {\sc xstar},
and consequently {\sc xillver}, automatically takes into account the
effects of Auger decay by accurately calculating the branching ratios
for Auger versus fluorescence emission; however, it does not include
resonant scattering of the lines.  This omission seems reasonable
because it is plausible that velocity gradients or turbulence in an
accretion disk will remove photons from the line core, thereby
reducing resonant scattering.  Nevertheless, because we disregard this
mechanism we may be over-predicting the line intensity for some
models.  As one additional complication, \cite{lie05b} has shown that
Auger resonance destruction is a selective process that highly
attenuates only a limited subset of Fe K$\alpha$ lines.  We conclude
that further analysis is required in order to accurately quantify the
emission from L-shell Fe ions.

\begin{figure}
\epsscale{1.0}\plotone{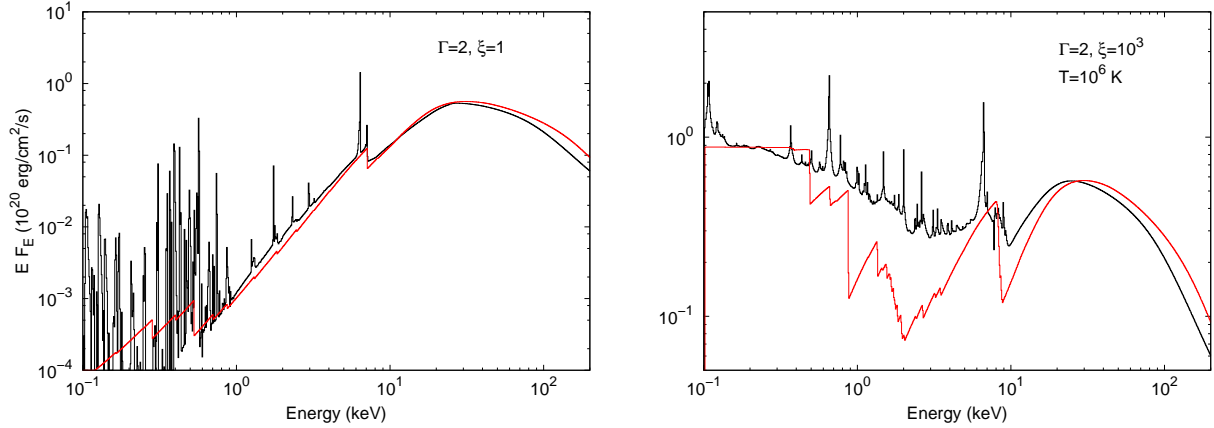}
\caption{Comparison of the reflected spectra calculated with {\sc xillver} (black curves),
and {\sc pexrav/pexriv} (red curves). Left panel shows the completely neutral case modeled
by {\sc pexrav}, compared with a {\sc xillver} model with $\xi=1$. Right panel shows the ionized
reflection case, thus a {\sc pexriv} model with $\xi=10^3$ is compared with {\sc xillver} at
the same ionization parameter. The disk temperature in the {\sc pexriv} model was set to
$T=10^6$~K (the highest possible value). In all cases $\Gamma=2$, the high-energy cutoff 
is at $300$~keV, and solar abundances are used.
}
\label{fpexrav}
\end{figure}

We have also compared our results to those obtained using the
reflection models {\sc pexrav} and {\sc pexriv} \citep{mag95}, which
compute the reflected spectrum from a completely neutral disk and an
ionized disk, respectively.  These models are an implementation of a
semi-analytic Green's function \citep{lig88} that models Compton
reflection of X-rays by cold electrons while including bound-free
continuum opacity.  Fluorescence emission line are however not
included.  The left panel of Figure~\ref{fpexrav} compares the
reflected spectrum predicted by {\sc xillver} with that predicted by
{\sc pexrav} for an incident power-law spectrum with $\Gamma=2$ and an
exponential cutoff at $300$~keV.  Because {\sc pexrav} models a
completely neutral gas, we use the lowest ionization parameter
considered in {\sc xillver}, namely $\xi=1$.  The two models agree
fairly well in the level of the reflected continuum, in particular
near the Fe K line region.  The edge energy and EW of the feature at
$\sim 7.2$~keV are similar.  Our model shows more absorption than {\sc
pexrav} in the low-energy part of the spectrum.  However, the {\sc
xillver} model can not be considered completely neutral; there is some
heating near the surface (see the leftmost curve in the middle panel
of Figure~\ref{ftemp.xi}).

The right panel in Figure~\ref{fpexrav} compares a reflected spectrum
from an ionized disk ($\xi=10^3$) computed using {\sc xillver} with
one computed using {\sc pexriv}.  At the outset, we note that {\sc
pexriv}, unlike both {\sc xillver} and {\sc reflionx}, does not solve
for the ionization balance of the slab.  Instead, {\sc pexriv} assumes
an isothermal medium with a maximum temperature of $T=10^6$~K.  Both
the ionization parameter and the gas temperature are allowed to vary.
This approach results in the large discrepancies that are evident in
both the level of the continuum and the edge features.  The level of
the continuum is similar only for energies below $1$~keV. For higher
energies, the {\sc pexriv} spectrum is highly absorbed and its Fe K
edge is much deeper.  This comparison demonstrates for the case of an
ionized reflector the importance of accurately modeling the ionization
balance.  {\sc pexriv}'s assumption of an isothermal gas is a poor one
for large values of the ionization parameter.  Another limitation of
{\sc pexriv} that was pointed out by \cite{fab10} is the lack of blurring
effects of Compton scattering, which smears the K edges.

We now discuss in turn the differences at high energies
($\gtrsim 20$~keV) between our results and those delivered by {\sc
reflionx} and {\sc pexrav/pexriv}.  At these energies, photoelectric
absorption is negligible, and the only source of opacity is Compton
scattering.  Furthermore, the gas temperature has only a very small
effect on the Comptonized spectrum.  Therefore, any differences in the
model spectra can be attributed to how each model treats scattering of
photons by cold electrons.  {\sc xillver} implements a Gaussian
convolution for all energies and at all depths (Section~\ref{secmod}).
{\sc reflionx} uses this approximation for the lines only, while using
a modified Kompaneets operator to treat the diffuse radiation.  

We tested the accuracy of our approximation by comparing with results
obtained using a Monte Carlo (MC) code, which we treat as the gold
standard.  This code, which extends the work of \cite{mat02} and
\cite{mag95}, simulates Compton scattering and radiative transfer
through neutral gas for a semi-infinite slab \citep{eik12}.  Included
in this code are photoabsorption for all elements with $Z \le 30$ and
fluorescent line emission using the fluorescent yields from
\cite{kas93}. Compton scattering is modeled using the proper
relativistic scattering dynamics and the differential Klein-Nishina
cross section.

\begin{figure}
\epsscale{1.0}\plotone{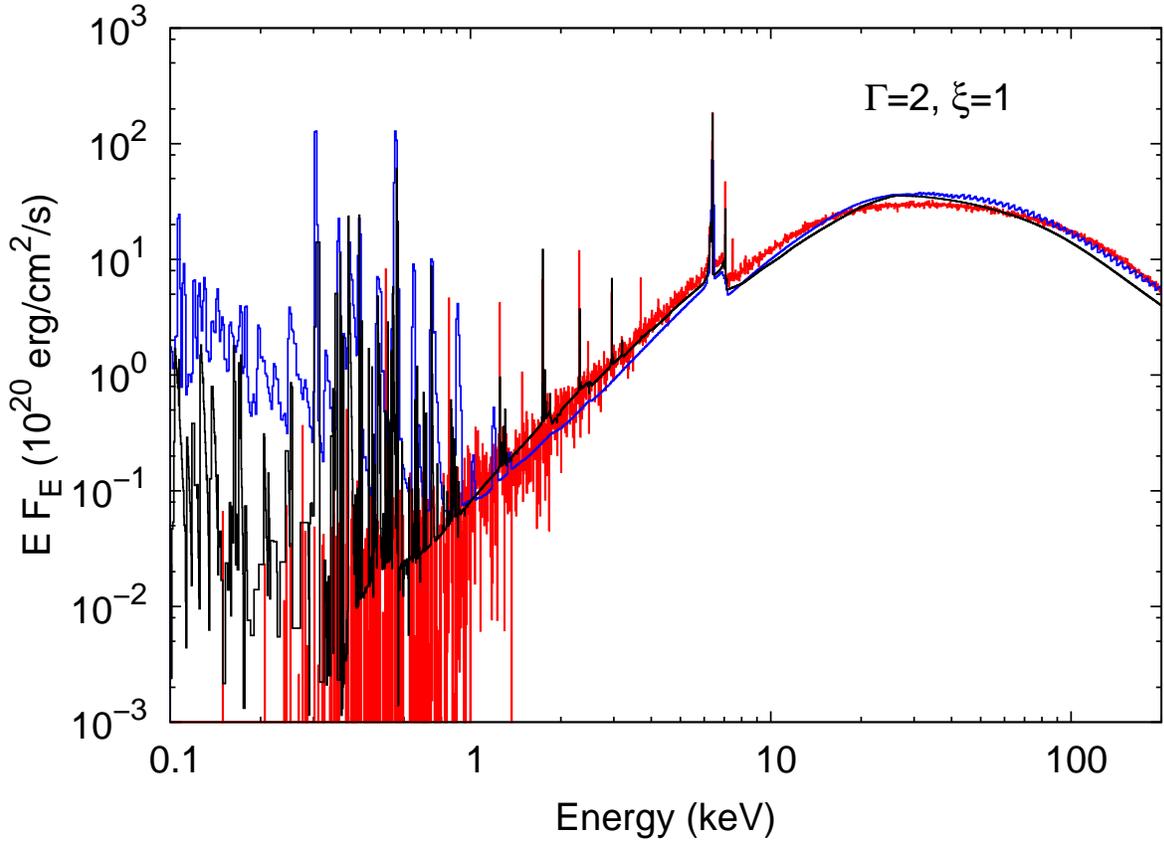}
\caption{Comparison of the reflected spectra as calculated with {\sc xillver} (black curve),
{\sc reflionx} (blue curve), and the Monte Carlo simulation (red curve), for an illumination
with $\Gamma=2$, $\xi=1$, and solar abundances.
}
\label{fmontecarlo}
\end{figure}

Figure~\ref{fmontecarlo} compares reflected spectra computed using
{\sc xillver} and the MC code for $\Gamma=2$, $\xi=1$ and solar
abundances.  The result for {\sc reflionx} is also shown for these same
parameters.  Overall, there is a good agreement between the three
models.  At energies $\lesssim 10$~keV, {\sc xillver} and the MC
simulation predict very nearly the same level of the continuum and the
same spectrum of intense emission lines.  For example, the $\sim
6.4$~keV Fe K$\alpha$ line has a very similar intensity and shape,
although the MC code predicts a $\sim 7.2$~keV K$\beta$ line that is
somewhat more intense.  The main differences occur above the $\sim
7.2$~keV Fe~K edge, in the continuum feature referred to as the
Compton hump.  The MC simulated spectrum has a somewhat shallower Fe K
edge than the other two spectra, and its $\sim 20 - 60$~keV is
somewhat fainter.  Above $\sim100$~keV, the {\sc
reflionx} spectrum is the better match to the MC spectrum.  

As a bottom line, the differences between the MC simulation and either
{\sc xillver} or {\sc reflionx} are not larger than $30\%$, which is
reasonable given the simplistic approximation used in those models.
However, this level of performance may not be adequate for the
analysis of such current and future X-ray missions as NuSTAR
\citep{har10} and eROSITA \citep{pre11}.  Furthermore, the
Compton-hump spectrum can provide important constraints on disk
inclination.  Therefore, our goal is to improve this aspect of the
code in future versions of {\sc xillver}.
%
%
\section{Conclusions}\label{seccon}
In this paper we have presented a new and complete library of
synthetic spectra for modeling radiation that is reprocessed in an
accretion disk and emitted as a reflected spectrum, which is generated
in response to illumination by an incident power-law spectrum.  This
present version of our code {\sc xillver} is an update of those
presented in \cite{gar10} and \cite{gar11}.  We have made several
improvements to both the routines and the atomic data and have
produced a large grid of reflection models covering a wide range of
parameters.  Each model is characterized by the photon index $\Gamma$
of the illuminating radiation, the ionization parameter $\xi$ at the
surface of the disk, and the Fe abundance with respect to its solar
value.  A total of 720 reflected spectra are provided in a single FITS
file\footnote{\url{http://hea-www.cfa.harvard.edu/~javier/xillver/}}
suitable for the analysis of X-rays observations via the {\tt
atable} model in {\sc xspec}. In order to represent the physical
conditions typically observed in most accreting sources, this library
covers the following range of parameters: $1.2 \leq \Gamma \leq 3.4$,
$1 \leq \xi \leq 10^4$, and $0.5 \leq A_\mathrm{Fe} \leq 10$.

The spectrum that illuminates the surface of the accretion disk is
assumed to be a power-law in the $10^{-1} - 10^6$~eV energy range,
with a sharp low-energy cutoff at $100$~eV and an exponential
high-energy cutoff at $300$~keV.  The low-energy cutoff is important
for spectra with steep power laws ($\Gamma > 2$); it prevents the
spectrum from being unphysically over-populated with low-energy
photons.  The power-law is extended up to $1$~MeV because in the
low-$\Gamma$ case this portion of the spectrum is rich in high-energy
photons, which importantly effect both the gas temperature and the
reflected spectrum.

The intensity of the illuminating spectrum, which is specified by the
ionization parameter $\xi$, significantly impacts the ionization
structure of the gas.  In all models, the gas temperature in the
vertical direction has a similar profile.  Near the surface, where the
illumination is intense, there is a hot layer ($T \gtrsim 10^6$~K)
where Compton heating and cooling dominate.  At larger depths, a
warm/cold region exists where photoelectric opacity and recombination
dominate.  The depth at which the transition between these two regimes
occurs increases as $\xi$ increase and as the incident spectrum
softens (large $\Gamma$).  Above the transition layer, the radiation
field thermalizes and the gas temperature remains fairly constant at
around $10^5$~K.  For low-ionization ($\xi \lesssim 100-500$), the
reflected spectrum displays a very rich set of emission lines
superimposed on a strongly absorbed continuum.  For high-ionization
($\xi \gtrsim 500-1000$), the spectrum consists of very narrow
emission lines from ionized species and a continuum that resembles the
incident power-law.

We have presented detailed comparisons with the reflection models {\sc
reflionx} \citep{ros05}, and {\sc pexrav} \citep{mag95}.  The {\sc
reflionx} model is more similar to ours because of its complexity and
the range of physical processes considered.  Furthermore, {\sc
reflionx} and {\sc xillver} both characterize the illuminating
spectrum as a power law, cover the same energy range, and use the same
definition for the input parameters.  The spectra generated by the two
codes are in good agreement.  The spectra agree best for large
$\Gamma$ (soft spectra) and low values of $\xi$.  However, at low
energies {\sc reflionx} generates an excess of soft flux compared to
{\sc xillver}.  For hard spectra, $\Gamma=1.4$, and high ionization,
$\xi=10^3$, the difference in the continuum at $1$~keV is about an
order of magnitude.  This large discrepancy is not well understood and
requires further analysis.

At very high or very low ionization, {\sc xillver} and {\sc reflionx}
predict similar EWs for the Fe K emission feature (including both
K$\alpha$ and K$\beta$), although the values for {\sc xillver} are
generally larger.  However, at intermediate levels of ionization,
in the $10^2 \lesssim \xi \lesssim 10^3$ range, the two models disagree
strongly: the Fe K EWs computed in this range computed using {\sc
reflionx} drop drastically.  The effect is presumably because 
{\sc reflionx} assumes that the Fe K lines from second-row ions are
completely suppressed by Auger resonant destruction.

Comparing {\sc xillver} with the more simplistic models {\sc pexrav}
and {\sc pexriv} highlights the importance of performing detailed
calculations which take into account key physical processes.  The
greatest deficiency of these models is that they do not generate
emission lines.  The neutral {\sc pexrav} model is in reasonable
agreement with {\sc xillver} concerning the level of the continuum and
the strength of the edges.  However, for moderate or high ionization
($\xi \gtrsim 20$), the {\sc pexriv} model is in strong disagreement
because it assumes an isothermal disk, as previously pointed out by
\cite{fab10}.

The library of models presented in this paper is suitable for modeling
the reflection spectra of accreting sources when the thermal disk
component of emission is small compared to the incident power-law
component.  That is, the present version of {\sc xillver} is suitable
for analyzing the spectra of AGN, and also GBHs in the hard state,
when the disk component is cool and faint.  For GBHs in the thermal or
steep power-law states \citep{rem06}, the strong
thermal component entering the atmosphere from below will
significantly change the ionization structure of the disk.  In a
future publication, we will report on an extension of {\sc xillver}
that is appropriate for modeling reflection spectra in the presence of
a strong thermal component.

Regardless of how accurate {\sc xillver} (or any such slab reflection
model) becomes, it is by itself inadequate for modeling the reflection
component in the spectra of accreting black holes.  For example,
current black hole spin determinations based on fitting of the Fe K
line naively assume a single ionization state for the reflecting
portion of the disk, which can extend over hundreds of gravitational
radii.  We plan to construct more realistic models for particular
illumination geometries that allow for an ionization gradient in the
radial direction.  This work will combine the general relativistic
approach implemented in the {\sc relline} code \citep{dau10,dau13}
with the model spectra presented in this paper.
%
\acknowledgments
%
%
%
\bibliographystyle{apj}
\bibliography{my-references}
%
%
%
\begin{deluxetable}{cccc}
\tablecaption{Elemental Solar abundances
\label{tabund}}
\tablewidth{0pt}
\tablehead{
\colhead{Element} & \colhead{{\sc xillver}\tablenotemark{a}} & \colhead{{\sc reflionx}\tablenotemark{b}} & \colhead{ Ratio} \\
}
\startdata
H   &  $1.0$                &  $1.0$               & $1.00$ \\
He  &  $0.1$                &  $0.1$               & $1.00$ \\
C   &  $3.7\times 10^{-4}$  &  $4.5\times 10^{-4}$ & $0.83$ \\
N   &  $1.1\times 10^{-4}$  &  $9.1\times 10^{-5}$ & $1.21$ \\
O   &  $6.8\times 10^{-4}$  &  $7.4\times 10^{-4}$ & $0.92$ \\
Ne  &  $2.8\times 10^{-5}$  &  $1.4\times 10^{-4}$ & $0.20$ \\
Mg  &  $3.5\times 10^{-5}$  &  $4.0\times 10^{-5}$ & $0.88$ \\
Si  &  $3.5\times 10^{-5}$  &  $3.7\times 10^{-5}$ & $0.94$ \\
S   &  $1.6\times 10^{-5}$  &  $1.9\times 10^{-5}$ & $0.84$ \\
Ar  &  $4.5\times 10^{-6}$  &  $3.8\times 10^{-6}$ & $1.18$ \\
Ca  &  $2.1\times 10^{-6}$  &  $2.2\times 10^{-6}$ & $0.94$ \\
Fe  &  $2.5\times 10^{-5}$  &  $3.3\times 10^{-5}$ & $0.76$ \\
\enddata
\tablenotetext{a}{\cite{gre98}}
\tablenotetext{b}{\cite{mor83}}
\end{deluxetable}
%

\newpage
%
%
\end{document}